\documentclass[11pt]{article}
\pdfoutput=1
\usepackage[utf8]{inputenc}
\usepackage{dsfont}
\usepackage{amsfonts}
\usepackage{amsmath}
\allowdisplaybreaks[4]        
\usepackage{amssymb}
\usepackage{euscript}     
\usepackage{braket}
\usepackage{starfont}
\usepackage{color,soul}         
\usepackage{tensor}        
\usepackage{amsthm}
\usepackage{graphicx}
\usepackage{slashed}
\usepackage{leftidx}
\usepackage{subfigure}
\usepackage{bbm}
\definecolor{outerspace}{rgb}{0.25, 0.29, 0.3}
\definecolor{scarlet}{rgb}{1.0, 0.13, 0.0}
\usepackage[header,title,page,titletoc]{appendix}  
\definecolor{princetonorange}{rgb}{1.0, 0.56, 0.0}
\definecolor{WildStrawberry}{rgb}{1.0, 0.26, 0.64}
\definecolor{rossocorsa}{rgb}{0.83, 0.0, 0.0}
\definecolor{navyblue}{rgb}{0.0, 0.0, 0.5}
\usepackage[numbers,sort&compress]{natbib}  
\usepackage{float}
\usepackage[paper=letterpaper,margin=1in]{geometry}
\newtheorem{theorem}{Theorem}

\newtheorem{defi}{Definition}

\newcommand{\req}[1]{(\ref{#1})} %{Eq.\thinspace(\ref{#1})}  

\newcommand{\bea}{\begin{eqnarray}}
\newcommand{\diff}{\mathrm{d}}
\newcommand{\eea}{\end{eqnarray}}
\newcommand{\ba}{\begin{eqnarray}}
\newcommand{\ea}{\end{eqnarray}}

\newcommand{\be}{\begin{equation}}
\newcommand{\ee}{\end{equation} }
\newcommand{\beqa}{\begin{eqnarray}}
\newcommand{\eeqa}{\end{eqnarray}}
\newcommand{\beqar}{\begin{eqnarray*}}
\newcommand{\eeqar}{\end{eqnarray*}}

\renewcommand{\req}[1]{eq.~(\ref{#1})}

\newcommand{\eg}{{\it e.g.,}\ }
\newcommand{\ie}{{\it i.e.,}\ }

\usepackage{hyperref}
\hypersetup{
    colorlinks,
    citecolor=rossocorsa,
    filecolor=navyblue,
    linkcolor=navyblue,
    urlcolor=navyblue
}

\begin{document} 

\begin{titlepage}

\begin{center}

\phantom{ }
\vspace{3cm}

{\bf \Large{(Generalized) quasi-topological gravities at all orders}}
\vskip 0.5cm
Pablo Bueno,${}^{\text{\Zeus}}$ Pablo A. Cano${}^{\text{\Kronos}}$ and Robie A. Hennigar${}^{\text{\Apollon}}$
\vskip 0.05in
\small{${}^{\text{\Zeus}}$ \textit{Instituto Balseiro, Centro At\'omico Bariloche}}
\vskip -.4cm
\small{\textit{ 8400-S.C. de Bariloche, R\'io Negro, Argentina}}

\small{${}^{\text{\Kronos}}$ \textit{Instituto de F\'isica Te\'orica UAM/CSIC}}
\vskip -.4cm
\small{\textit{ C/ Nicol\'as Cabrera, 13-15, C.U. Cantoblanco, 28049 Madrid, Spain}}

\small{${}^{\text{\Apollon}}$ \textit{Department of Mathematics and Statistics, Memorial University of Newfoundland}}
\vskip -.4cm
\small{\textit{  St. John's, Newfoundland and Labrador, A1C 5S7, Canada}}

%\vskip -.4cm
%\small{\textit{}}
\begin{abstract}

%Generalized quasi-topological gravities are higher-curvature extensions of $D$-dimensional Einstein gravity

%Generalized quasi-topological gravities (GQTGs) are higher-curvature ex-tensions of Einstein gravity characterized by the existence of non-hairy generalizationsof the Schwarzschild black hole which satisfygttgrr=−1, as well as for having second-order  linearized  equations  around  maximally  symmetric  backgrounds.

%Quasi-topological gravities are higher-curvature modifications of $D$-dimensional Einstein gravity characterized by possessing non-hairy generalizations of the Schwarzschild black hole satisfying $g_{tt}g_{rr}=-1$ and by having second-order equations of motion when linearized around maximally symmetric backgrounds.

A new class of higher-curvature modifications of $D(\geq 4$)-dimensional Einstein gravity has been recently identified. Densities belonging to this ``Generalized quasi-topological''  class (GQTGs) are characterized by possessing non-hairy generalizations of the Schwarzschild black hole satisfying $g_{tt}g_{rr}=-1$ and by having second-order equations of motion when linearized around maximally symmetric backgrounds.  GQTGs for which the equation of the metric function $f(r)\equiv -g_{tt}$ is algebraic are called ``Quasi-topological'' and only exist for $D\geq 5$. In this paper we prove that GQTG and Quasi-topological densities exist in general dimensions and at arbitrarily high curvature orders. We present recursive formulas which allow for the systematic construction of $n$-th order densities of both types from lower order ones, as well as explicit expressions valid at any order. We also obtain the equation satisfied by $f(r)$ for general $D$ and $n$.  Our results here tie up the remaining loose end in the proof presented in {\tt arXiv:1906.00987} that every gravitational effective action constructed from arbitrary contractions of the metric and the Riemann tensor is equivalent, through a metric redefinition, to some GQTG.

\end{abstract}
\end{center}
\end{titlepage}

\setcounter{tocdepth}{2}

{\parskip = .2\baselineskip \tableofcontents}

%\maketitle
%\flushbottom
%%%%%%%%%%%%%%%%%%%%%%%%%%%%%%%%%%%%%
%\setcounter{tocdepth}{2}
%the line above sets the depth of the table of contents. {2} means it will display section and subsections only.
%{\small
%\setlength\parskip{-0.5mm} 
%\tableofcontents
%}

%%%%%%%%%%%%%%%%%%%%%%%%%%%%%%
\section{Introduction}
\label{sec:Introduction}
In recent times, a new class of higher-curvature modifications of the Einstein-Hilbert action has been identified. These so-called ``Generalized quasi-topological gravities'' (GQTGs) are covariant metric theories whose action can be generically written as  
  \begin{equation}\label{action}
  S=\frac{1}{16\pi G} \int \diff^D x \sqrt{|g|} \left[-2\Lambda+R+\sum_{n=2}\sum_{i_n} \ell^{2(n-1)} \mu_{i_n}^{(n)} \mathcal{R}_{i_n}^{(n)} \right]\, ,
  \end{equation}
 where  $\mathcal{R}_{i_n}^{(n)}$ are densities constructed from $n$ Riemann tensors and the metric,\footnote{In principle, GQTG densities involving covariant derivatives of the metric could also exist, but these have not been constructed to date.} the $\mu_{i_n}$ are dimensionless couplings, $\ell$ is some length scale, and $i_n$ is an index running over all independent GQTG invariants of order $n$. In general, GQTGs have fourth-order equations of motion. However, their defining property corresponds to the fact that, when evaluated on a general static and spherically symmetric ansatz ---see \req{SSS} below--- their equations of motion become second-oder, so they admit black holes characterized by the condition $g_{tt}g_{rr}=-1$ in Schwarzschild-like coordinates, \ie they take the form
%\comment{many refs missing}
\begin{equation}\label{fEq}
\diff s^2_{f}=-f(r)\diff t^2+\frac{\diff r^2}{f(r)}+r^2\diff \Omega^2_{(D-2)}\, ,
\end{equation}
where $f(r)$ satisfies an equation which contains at most two derivatives of such function ---so that \req{fEq} reduces to the usual Schwarzschild solution when we turn off all the $\mu_{i_n}$.  A particularly important subclass of GQTGs corresponds to Lovelock theories \cite{Lovelock1,Lovelock2}, which are the most general theories for which the second-order-equations condition actually holds for any metric. Besides Einstein gravity itself, there are no non-trivial Lovelock densities in $D=4$ ---the first non-trivial one in $D\geq 5$ corresponds to the Gauss-Bonnet density. Interestingly, non-trivial GQTGs do exist for $n\geq 3$ in $D=4$.

GQTG densities possess a series of interesting properties which have been studied in many papers \cite{PabloPablo,Hennigar:2016gkm,PabloPablo2,Hennigar:2017ego,PabloPablo3,Ahmed:2017jod,PabloPablo4,Dey:2016pei,Feng:2017tev,Hennigar:2017umz,Hennigar:2018hza,Bueno:2018xqc,Bueno:2018yzo,Bueno:2018uoy,Poshteh:2018wqy,Mir:2019ecg,Mir:2019rik,Arciniega:2018fxj,Cisterna:2018tgx,Arciniega:2018tnn,Mehdizadeh:2019qvc,Erices:2019mkd,Emond:2019crr,Jiang:2019fpz} and appear summarized in some detail \eg in \cite{Bueno:2019ltp}. Among the most relevant ones, we can mention: i) when linearized around any maximally symmetric background, their equations are identical to the Einstein gravity ones, up to a redefinition of the Newton constant ---in other words, they only propagate the usual transverse and traceless graviton in the vacuum \cite{PabloPablo,Hennigar:2016gkm,PabloPablo2,Hennigar:2017ego,PabloPablo3,Ahmed:2017jod,PabloPablo4};\footnote{Higher-curvature gravities satisfying this property ---but not necessarily the GQTG condition \req{GQTGcond}--- have been also studied in several other papers, \eg \cite{Aspects,Tekin1,Tekin2,Tekin3,Love,Ghodsi:2017iee,Li:2017ncu,Li:2017txk,Li:2018drw,Li:2019auk,Lu:2019urr}.} ii) they possess non-hairy black hole solutions fully characterized by their ADM mass/energy and whose thermodynamic properties can be obtained from an algebraic system of equations; iii) at least in $D=4$, black holes generically become thermodynamically stable below certain mass \cite{PabloPablo4}; iv) in addition to black holes, certain subsets of GQTGs also contain Taub-NUT/Bolt solutions characterized by a single metric function and analytic thermodynamics \cite{Bueno:2018uoy}; v) when evaluated on a Friedmann-Lema\^itre-Robertson-Walker (FLRW) ansatz, certain GQTGs in $D=4$ also give rise to second-order equations for the scale factor, with intriguing consequences regarding cosmological evolution \cite{Arciniega:2018fxj,Cisterna:2018tgx,Arciniega:2018tnn};
vi) we can consider arbitrary linear combinations of GQTG densities and the corresponding properties hold, which means, in particular, that GQTG theories have a well-defined and continuous Einstein gravity limit, corresponding to setting all higher-curvature couplings to zero.

In addition to these properties, in \cite{Bueno:2019ltp} it was shown that any higher-curvature effective action involving arbitrary contractions of the metric and the Riemann tensor can be mapped, via a metric redefinition, to some GQTG. This claim was shown to be true as long as at least one GQTG density exists at every order in curvature. Here we complete this proof by explicitly constructing GQTG densities for general $n$ and $D$.

Among all GQTG densities, special mention deserve the so-called Quasi-topological theories \cite{Quasi2,Quasi,Myers:2010jv}. For those, the equation characterizing the metric function $f(r)$ is algebraic and, at least in the cases studied, they satisfy a Birkhoff theorem \cite{Oliva:2011xu,Oliva:2012zs,Quasi2,Cisterna:2017umf}. Theories of this subclass only exist for $D\geq 5$. Prior to this paper, Quasi-topological gravities had been constructed order by order in curvature for $n=3$ \cite{Quasi2,Quasi}, $n=4$ \cite{Dehghani:2013ldu,Ahmed:2017jod} and $n=5$ in $D=5$ \cite{Cisterna:2017umf}. A particular subclass of Quasi-topological gravities ---characterized by the additional property that the trace of the field equations is second-order in derivatives of the metric for general backgrounds--- had been previously identified in $D=2n-1$ for general $n$ \cite{Quasi2,Oliva:2010zd,Oliva:2011xu}. Here, we conclude the order-by-order program, by providing explicit constructions of Quasi-topological densities of arbitrary curvature orders and in general dimensions.
%\footnote{In fact, the corresponding density for $n=2$ and $D=3$ turns out to correspond to the quadratic piece of the }

The remainder of the paper goes as follows. In Section \ref{GQTGss} we review the defining properties of GQTGs and (a particular subset of those corresponding to the so-called) Quasi-topological gravities. In Section \ref{recrel} we show how both GQTGs and Quasi-topological densities of arbitrary orders in curvature can be constructed recursively starting from lower-order densities. In Section \ref{explifo} we provide explicit formulas for $n$-th order GQTG and Quasi-topological densities. In Section \ref{feqss} we obtain the equations satisfied by the metric function $f(r)$ characterizing the static black holes of GQTG and Quasi-topological gravities. We conclude in Section \ref{discuss} with some prospects for future research.

%A curvature invariant of order $n$, $\mathcal{L}_{(n)}$, is said to be of the GQTG class when it satisfies the following property.
%The technical requirement which makes a generic $\mathcal{L}(g^{ab},R_{abcd},\nabla_a R_{bcde},\dots)$ theory %belong to the GQTG class is the following. 

\section{Generalized quasi-topological gravities}\label{GQTGss}
In this section we review the defining properties of GQTGs and introduce some terminology required for a proper understanding of the following sections. Special emphasis is put in the difference between the subclass of theories commonly known as ``Quasi-topological gravities'' and those GQTGs not belonging to this subset. %Among other things, we present a simple condition which distinguishes between both kinds of theories. 

\subsection{Definitions}
Let us start our discussion with a general static and spherically symmetric ansatz (SSS),
\begin{equation}
\label{SSS}
\diff s^2_{N,f}=-N(r)^2 f(r)\diff t^2+\frac{\diff r^2}{f(r)}+r^2\diff \Omega^2_{(D-2)}\, .
\end{equation}
Given some curvature invariant of order $n$, $\mathcal{R}_{(n)}$, consider the effective Lagrangian resulting from the evaluation of $\sqrt{|g|}\mathcal{R}_{(n)}$ in \req{SSS}
%and let $L_{N,f}$ be the effective Lagrangian which results from evaluating $\sqrt{|g|}\mathcal{L}$ in \req{SSS}, namely
\begin{equation}\label{ansS}
L_{N,f}\equiv \left. N(r) r^{D-2} \mathcal{R}_{(n)}\right|_{N,f}\, ,
\end{equation}
(up to an irrelevant angular contribution). The associated action evaluated in \req{SSS} reads
\be
S_{N,f}\equiv \left. \int \diff^Dx \sqrt{|g|}\,\mathcal{R}_{(n)} \right|_{N,f}\equiv \Omega_{(D-2)}\int \diff t  \int \diff r L_{N,f} \, ,
\ee
where $\Omega_{(D-2)}\equiv 2\pi^{\frac{(D-1)}{2}}/\Gamma[\frac{D-1}{2}]$.
Given a generic theory, imposing the full nonlinear equations of motion to be satisfied for a metric of the form (\ref{SSS}) can be seen to be equivalent to imposing the Euler-Lagrange equations associated to $N(r)$ and $f(r)$ ---see \eg \cite{Palais:1979rca,Deser:2003up,PabloPablo4}---,
\be
\left.\mathcal{E}^{ab}\right|_{N,f}\equiv \left. \frac{1}{\sqrt{|g|}} \frac{\delta S}{\delta g^{ab}} \right|_{N,f}=0 \quad  \Leftrightarrow \quad \frac{\delta S_{N,f}}{\delta N}=\frac{\delta S_{N,f}}{\delta f}=0\, .
\ee
Now, let $L_f\equiv L_{1,f}$ and $S_f\equiv S_{1,f}$, namely, the expressions resulting from setting $N=1$ in $L_{N,f}$. 
\begin{defi}
We say that $\mathcal{R}_{(n)}$ is of the GQTG class if the Euler-Lagrange equation of $f(r)$ associated to $L_f$ vanishes identically, \ie if\footnote{Naturally, the condition is exactly the same if, instead of a single density $\mathcal{R}_{(n)}$, we consider a full higher-order Lagrangian $\mathcal{L}(g^{ab},R_{abcd},\nabla_a R_{bcde},\dots)$ involving any linear combination of densities of arbitrary orders, as in \req{action}.}\footnote{In terms of $L_f$, this condition reads \begin{equation} \frac{\partial L_f}{\partial f}-\frac{\diff}{\diff r} \frac{\partial L_f}{\partial f'}+\frac{\diff^2}{\diff r^2}  \frac{\partial L_f}{\partial f''}=0\, , \quad \forall \, \, f(r)\, .\end{equation}}
%\footnote{Observe the slight abuse of notation in writing ``$\delta L_f/\delta f$''. This functional variation is in fact taken with respect to the action $S_{N,f}\equiv \Omega_{(D-2)}\int dt  \int dr L_{N,f}$. In order to minimize the clutter of the already messy equations, we }
\begin{equation}\label{GQTGcond}
\frac{\delta S_f}{\delta f} = 0\, , \quad \forall \, \, f(r)
\, .
\end{equation}
This is equivalent to $L_f$ being a total derivative, namely, to
\begin{equation}\label{condd2}
L_f =F_0'\, ,
\end{equation}
for some function $F_0(r,f(r),f'(r))$.
\end{defi}
%In that case, the explicit form of  corresponding density allows for a solution of the form
%Hence, while we need to consider the general ansatz \req{SSS} to obtain $f(r)$ for a given theory, 

Observe that the above conditions exclusively depend on the on-shell Lagrangian evaluated on the single-function ansatz appearing in \req{SSS} with $N=1$, \ie on \req{fEq}. However, in order to obtain the equation satisfied by $f(r)$ for a given GQTG density from this on-shell Lagrangian method, we need to compute $L_{N,f}$. Then, the equation of $f(r)$ can be obtained from its variation with respect to $N(r)$, \ie
\begin{equation}\label{eqf}
\left.\frac{\delta S_{N,f}}{\delta N}\right|_{N=1}=0\, \quad \Leftrightarrow\quad \text{equation for}\quad f(r)\, .
\end{equation}
As argued in \cite{PabloPablo3}, provided \req{condd2} holds, the effective Lagrangian $L_{N,f}$ always takes the form
\begin{equation}\label{fofwo}
L_{N,f}=N F_0' + N' F_1 + N'' F_2 +\mathcal{O}(N'^2/N)\, ,
\end{equation}
where $F_{1}$ and $F_2$ are functions of $f(r)$ and its derivatives, and $\mathcal{O}(N'^2/N)$ is a sum of terms which are at least quadratic in derivatives of $N(r)$. Then, integrating by parts, it can be shown that 
\begin{equation}
S_{N,f}=\Omega_{(D-2)}\int dt \int dr \left[N\left(F_0-F_1+F_2' \right)' +\mathcal{O}(N'^2/N) \right]\, .
\end{equation}
The structure is such that we can write all terms involving a single power of $N(r)$ (or its derivatives) as a product of $N(r)$ and a total derivative which is a function of $f(r)$ alone. Imposing condition (\ref{eqf}) equates that total derivative piece to zero, so we can integrate it once, yielding \cite{PabloPablo3}
\begin{equation} \label{eqqqf}
\mathcal{F}_{\mathcal{R}_{(n)}}  \equiv F_0-F_1+F_2'=C\, ,
\end{equation}
where the integration constant $C$ will be related to the ADM mass of the solution \cite{Arnowitt:1960es,Arnowitt:1960zzc,Arnowitt:1961zz,Deser:2002jk}. Hence, given some GQTG density $\mathcal{R}_{(n)}$, we just need to evaluate $L_{N,f}$ as defined in \req{ansS} and then identify the functions $F_{i=0,1,2}$ from \req{fofwo}. The equation of $f(r)$ is then given by \req{eqqqf}.

Since we have integrated once, one would naively expect this equation to be third-order in derivatives of $f(r)$ in general ---see \req{eomGHOG} below. However, as explained in \cite{PabloPablo3}, the $b=r$ component of the Bianchi identity, $\nabla_a \mathcal{E}^{ab}=0$, relates ${\diff \mathcal{E}^{rr}/ \diff r} $ to the rest of nonvanishing components (without derivatives), which implies that $ \mathcal{E}^{rr}$ is in fact third-order in derivatives of $f(r)$. It follows that \req{eqqqf}, where we have integrated once, is at most second-order in derivatives of $f(r)$.  In the following subsection we will subdivide GQTGs in two groups: those for which \req{eqqqf} indeed involves $f'(r)$ and $f''(r)$, and those for which \req{eqqqf} is in fact an algebraic equation. We analyze these two possibilities in more detail in the following subsection.
The explicit form of the equation will be obtained in Section \ref{feqss} for both types of theories for general $D$ and $n$.

\subsection{Lovelock vs Quasi-topological vs Generalized quasi-topological}
\label{secLoveQ}

For a general higher-curvature density constructed from arbitrary contractions of the Riemann tensor and the metric, the nonlinear equations of motion involve up to four derivatives of $g_{ab}$ and can be written generically as (see \eg \cite{Padmanabhan:2013xyr})
\begin{equation}\label{eomGHOG}
\mathcal{E}_{ab}\equiv P_{a}\,^{cde}R_{bcde}-\frac{1}{2}g_{ab}\mathcal{R}_{(n)}-2\nabla^{e}\nabla^{f}P_{aefb}=0\, , \quad \text{where} \quad P^{abcd}\equiv \left[\frac{\partial \mathcal{R}_{(n)}}{\partial R_{ab dc}}\right]_{g^{ef}}\, .
\end{equation}
All contributions involving more than two derivatives come from the last term in $\mathcal{E}_{ab}$. 
The class of theories for which such term is absent are the well-known Lovelock gravities \cite{Lovelock1,Lovelock2}, which are indeed the most general covariant metric theories with second-order equations of motion. For a given $n$, the Lovelock density $\mathcal{X}_{(n)}$ turns out to be defined in general $D$ so that it reduces to the Euler density of compact manifolds when considered in $D=2n$. As a consequence of the above condition, Lovelock theories satisfy \req{GQTGcond}, so they are the simplest examples of GQTGs. When considered for a static and spherically symmetric ansatz, $\mathcal{E}^{rr}$ is first-order in derivatives of $f(r)$ and therefore \req{eqqqf} is just an algebraic (polynomial) equation for $f(r)$ \cite{Wheeler:1985nh,Wheeler:1985qd,Boulware:1985wk,Cai:2001dz,Dehghani:2009zzb,deBoer:2009gx,Camanho:2011rj,Garraffo:2008hu}

The next-to-simplest case corresponds to the so-called Quasi-topological gravities \cite{Quasi,Quasi2,Myers:2010jv,Dehghani:2013ldu,Cisterna:2017umf}. As opposed to Lovelock theories, Quasi-topological gravities have fourth-order equations of motion on general backgrounds. However, they behave very similarly to Lovelock theories when considered on solutions of the form (\ref{SSS}). In particular, their distinguishing property is that \req{eqqqf} is also algebraic. However, it has also been shown that all Quasi-topological gravities constructed to date satisfy a Birkhoff theorem when the background is promoted to a time-dependent one \cite{Oliva:2011xu,Oliva:2012zs,Quasi2,Cisterna:2017umf}.

% \comment{time-dependence comments here and below probably worth mentioning}. 

\begin{defi}
We say that a GQTG density belongs to the Quasi-topological subclass if the equation of $f(r)$ ---\req{eqqqf}--- is algebraic, \ie if it does not involve derivatives of $f(r)$.
\end{defi}

From this perspective, Lovelock theories are in turn a subset of the Quasi-topological subclass. Just like for them, $\mathcal{E}^{rr}$ is first-order for Quasi-topological gravities, which is a reflection of the fact that
\begin{equation}\label{qtt}
\nabla^a P_{abcd}|_f=0 \, ,
\end{equation}
\ie the piece responsible for the higher-derivative terms vanishes when evaluated on the static and spherically symmetric metric \req{fEq}. However, let us note that Quasi-topological theories appear to follow from the apparently less-restrictive condition of demanding that the trace of the field equations is proportional to the Lagrangian itself ---\ie $\nabla^a P_{a b}|_f = 0$ for $P_{ab}|_f = P^c{}_{a c b}|_f$.  Then, given some density $\mathcal{Z}_{(n)}$ satisfying the GQTG condition \req{GQTGcond} ---or, equivalently \req{condd2}---, it is possible to show that this will be of the Quasi-topological subclass if 
\be  \label{qtg22}
\left[ \frac{(D-2)}{r} \frac{\partial}{\partial f''} + \frac{\diff}{\diff r} \frac{\partial }{\partial f''} + \frac{(D-3)}{2} \frac{\partial }{\partial f'}  + \frac{r}{2} \frac{\diff}{\diff r} \frac{\partial}{\partial f'} - r \frac{\partial }{\partial f}\right]  \mathcal{Z}_{(n)}|_f = 0\, ,
\ee
which follows from the condition of second-order traced field equations for SSS backgrounds ---see Section~\ref{feqss} where we will make use of the explicit form of $P_{abcd}$. It would be interesting to determine if these conditions leading to algebraic field equations for the SSS backgrounds also imply a Birkhoff theorem for time-dependent metric ansatze.

%\comment{perhaps explain in some detail how it actually follows}

In sum, ``Quasi-topological'' gravities are a subset of the broader GQTG class which includes, in particular, Lovelock theories. Such subset can be characterized by condition \req{qtg22}. In this paper, we will distinguish between Quasi-topological gravities ---which we will denote by $\mathcal{Z}_{(n)}$--- and non-trivial GQTGs not belonging to the Quasi-topological class ---which we will denote by $\mathcal{S}_{(n)}$. For simplicity, we will simply refer to the latter as GQTGs from now on. While GQTGs exist in $D=4$ for $n\geq 3$, the only nontrivial Quasi-topological (and Lovelock) theory in $D=4$ is Einstein gravity. For $D\geq 5$, on the other hand, Quasi-topological and GQTG densities coexist for general curvature orders. Note finally that there exists yet another subset of GQTGs, which consists of densities that become trivial when evaluated on \req{SSS}.

\section{Recursive relations}\label{recrel}
In this section we show that Quasi-topological and GQTGs of arbitrary curvature orders  can be constructed from  simple recursive relations involving lower-order densities of the same kind (in each case). This concludes the program of constructing examples of this kind of theories order by order in curvature. It also completes the proof of Ref. \cite{Bueno:2019ltp} that any higher-curvature effective action involving arbitrary contractions of the Riemann tensor and the metric can be written, via metric redefinitions, as a GQTGs. 

\subsection{Generalized quasi-topological gravities}
For a metric of the form (\ref{fEq}), %---or, more generally, for one with horizon curvature $k=1,0,-1$ in the spherical, planar and hyperbolic cases respectively --- 
 the Riemann tensor can be conveniently written as \cite{Deser:2005pc}
\begin{equation}\label{rieef}
\left.\tensor{R}{^{ab}_{cd}}\right|_f=2\left[-A T^{[a}_{[c}T^{b]}_{d]}+2B T^{[a}_{[c}\sigma^{b]}_{d]}+\psi \sigma^{[a}_{[c}\sigma^{b]}_{d]}\right]\, ,
\end{equation}
where $T_a^b$ and $\sigma_a^b$ are, respectively, projectors on the ($t$,$r$) and angular directions,\footnote{They satisfy: $\sigma_a^b \sigma_b^c=\sigma_a^c$, $T_a^b T_b^c=T_a^c$, $\sigma_a^b T_b^c=0$, $\delta^a_b \sigma_a^b=(D-2)$, $\delta^a_b T_{a}^b=2$, $\delta^{a}_{b}=T^{a}_{b}+\sigma^{a}_{b}$. }
and we defined the functions $A$, $B$ and $\psi$ as
\begin{equation}
A\equiv \frac{f''(r)}{2}\, ,\quad B\equiv -\frac{f'(r)}{2r}\, ,\quad \psi\equiv \frac{k-f(r)}{r^2}\, ,
\end{equation}
where we have relaxed the spherical-horizon condition, the constant $k$ taking the values ${1,0, -1}$ for spherical, planar, and hyperbolic horizons, respectively.

Now, examples of GQTGs have been constructed in general dimensions and for considerably high orders of curvature \cite{PabloPablo,Hennigar:2016gkm,PabloPablo2,Hennigar:2017ego,PabloPablo3,Ahmed:2017jod,PabloPablo4,Arciniega:2018tnn}. Analyzing the existent cases, we observe that all of them satisfy the total derivative condition (\ref{condd2}) with 
\be  \label{totit}
r^{D-2} \mathcal{S}_{(n)}|_{f} =   \frac{\diff}{\diff r} \bigg[2 (n-2)  r^{D-1} \left( B +  \frac{(D-4)}{4}\psi \right)^{n-1} \left( B  -  \frac{(D-4 + 2n)}{2(n-2)}\psi\right)   \bigg]\, .
\ee
This expands to 
\begin{align} \notag
\mathcal{S}_{(n)}|_{f}  = + \frac{1}{2} &\bigg[(n-1) \Big((D-4) \psi - (D-6) B + 2 A \Big) \Big( (D-4 + 2n)\psi - 2(n-2)B \Big) \\ \label{densiexp}
&- 2\left(B + \frac{(D-4)}{4} \psi \right) \Big((D-3)(D-4 + 2n) \psi \\ \notag & + \left((6-2n)D + 8(n-2) \right) B + 2(n-2)A \Big) \bigg] \left(B + \frac{(D-4)}{4} \psi \right)^{n-2}\, .
\end{align}
Observe that the overall factor in the rhs of \req{totit} and \req{densiexp} is not fixed, as it can be modified by redefining the corresponding gravitational coupling. For instance, for $\mathcal{S}_{(m)}$, $m=1,2,3,4,5$, we may choose
\begin{align}\label{eg}
\mathcal{S}_{(1)}=&-R\, ,\\ 
\mathcal{S}_{(2)}=&-\frac{D}{4(D-2)(D-3)} \left[R^2-4R_{ab}R^{ab}+R_{abcd}R^{abcd} \right]\, , \\
\mathcal{S}_{(3)}=& +\frac{3 D R R_{ab} R^{ab} }{4 (D-1)(D-2)^2} - \frac{(D^2 + 8 D - 8) R^3}{16(D-1)^2(D-2)^2} - \frac{3 R^{ab} R^{cd} W_{acbd}}{2(D-2)(D-3)}\nonumber\\
& - \frac{3D R W_{abcd}W^{abcd}}{16(D-1)(D-2)(D-3)} 
+ \frac{(D-2)(2D-1) W_{ab}{}^{cd} W_{cd}{}^{ef} W_{ef}{}^{ab}}{8 (D-3) (D^3 - 9 D^2 + 26 D - 22) }\, ,
% +14\tensor{R}{_{a}^{c}_{b}^{d}}\tensor{R}{_{c}^{e}_{d}^{f}}\tensor{R}{_{e}^{a}_{f}^{b}} +2\tensor{R}{_{a bcd}}\tensor{R}{^{a bc}_{e}}R^{d e}-\frac{(38-29 D+4 D^2)}{4(D-2)(2D-1)}\tensor{R}{_{abcd}}\tensor{R}{^{abcd}}R\\
%&-\frac{2(-30+9D+4D^2)}{(D-2)(2D-1)}\tensor{R}{_{abcd}}\tensor{R}{^{ac}}\tensor{R}{^{bd}}
%-\frac{4(66-35D+2D^2))}{3(D-2)(2D-1)}R^{b}_{a} R_{b}^{c} R_{c}^{a}\\
%&+\frac{(34-21D+4D^2)}{(D-2)(2D-1)}R_{ab}R^{ab} R-\frac{(30-13D+4D^2)}{12(D-2)(2D-1)} R^3\, ,
\end{align}
while ${\cal S}_{(4)}$ and ${\cal S}_{(5)}$ are presented in the appendix. Note these expressions are normalized in a way such that  \req{totit} and \req{densiexp} hold as they are.

While the pattern in \req{totit} is suggestively simple, there is a priori no guarantee that such expression will hold for arbitrarily high curvature densities. However, a careful inspection of \req{totit} reveals the existence of an interesting recursive relation between densities of different orders satisfying such condition, namely,
\begin{align}\label{recu1}
\mathcal{S}_{(n + 5)} = -\frac{3(n+3)\mathcal{S}_{(1)} \mathcal{S}_{(n+4)} }{4(D-1)(n+1)}  + \frac{3 (n+4)\mathcal{S}_{(2)} \mathcal{S}_{(n+3)} }{4 (D-1)n}  
- \frac{(n+3)(n+4)\mathcal{S}_{(3)} \mathcal{S}_{(n+2)}}{4(D-1)n(n+1)}   \, .
\end{align}
It is not difficult to verify that replacing the explicit expressions of $\mathcal{S}_{(m)}|_f$ with $m=1,2,3,(n+2),(n+3),(n+4)$ shown above in the rhs, one is left with an expression which precisely agrees with the one corresponding to $\mathcal{S}_{(n+5)}|_f$. This means that starting from the first five densities we can construct densities of arbitrarily high order in curvature by iteratively applying \req{recu1}. Such densities will automatically be of the GQTG class by construction. The existence of relation \req{recu1} automatically proves that GQTG densities exist at all orders in curvature.

%Now let us introduce the following tensors
%\begin{align}
%\tensor{K}{^{ab}_{cd}}=&\tensor{R}{^{ab}_{cd}}-4 R^{[a}_{[c}\delta^{b]}_{d]}+R\delta^{[a}_{[c}\delta^{b]}_{d]}\, ,\\
%\tensor{Q}{^{ab}_{cd}}=&\tensor{R}{^{ab}_{cd}}-2 R^{[a}_{[c}\delta^{b]}_{d]}\, .
%\end{align}

%They have the following form when evaluated on the single-function metric
%\begin{align}
%\tensor{K}{^{ab}_{cd}}=&+(D-3)(D-2)\psi T^{[a}_{[c}T^{b]}_{d]} +\left[4B(D-3)+2(D-4)(D-3)\psi\right] T^{[a}_{[c}%\sigma^{b]}_{d]} \\ \notag & +\left[4B(D-4)-2A+(D-5)(D-4)\psi\right]\sigma^{[a}_{[c}\sigma^{b]}_{d]}\, ,\\
%\tensor{Q}{^{ab}_{cd}}=&-2B(D-2) T^{[a}_{[c}T^{b]}_{d]}+\left[2A-2(D-3)\psi-2B(D-2)\right] T^{[a}_{[c}\sigma^{b]}_{d]} \\ \notag &- (4B+2(D-4)\psi) \sigma^{[a}_{[c}\sigma^{b]}_{d]}\, .
%\end{align}

\subsection{Quasi-topological gravities}

Similarly to the case of the GQTGs densities analyzed above, we observe that all known Quasi-topological densities can be normalized so that, when evaluated in \req{fEq}, they can be written as
\be  \label{totaldQT}
\left. r^{D-2}{\cal Z}_{(n)}\right|_{f} = \frac{\diff}{\diff r} \left[r^{D-1} \psi^{n-1} \left((2n-D)\psi - 2 n B \right) \right] \, ,
\ee
which expands to
\be  \label{sifj}
\left.{\cal Z}_{(n)}\right|_f = -4n (n-1) B^2 \psi^{n-2} +  n\left(2 A- 4 (D-2n) B\right) \psi^{n-1} - (D-2n)(D-2n-1) \psi^n\, .
\ee

For ${\cal Z}_{m}$, $m=1,2,3,4,5$, we can choose
\begin{align} \label{z1}
{\cal Z}_{(1)} =& - R\, ,
\\
{\cal Z}_{(2)} =& - \frac{1}{(D-2)(D-3)} \left[R^2- 4 R_{ab}R^{ab}+R_{abcd}R^{abcd}   \right]\, ,
\\
{\cal Z}_{(3)} =& - \frac{8 (2D-3)}{(D-2)(D-3)(D - 4)(3D^2 - 15D + 16)} \Bigg[ (D - 4){{{R_a}^b}_c{}^d} {{{R_b}^e}_d}^f {{{R_e}^a}_f}^c \nonumber  \\ \label{ZD} 
              &  +  \frac{3(3D - 8)}{8(2D - 3)} R_{a b c d} R^{a b c d} R  - \frac{3(3D-4)}{2(2D - 3)} {R_a}^c {R_c}^a R  \\ \notag &- \frac{3(D-2)}{(2D - 3)} R_{a c b d} {R^{a c b}}_e R^{d e} + \frac{3D}{(2D - 3)} R_{a c b d} R^{a b} R^{c d} \\ \nonumber
              & 
                + \frac{6(D-2)}{(2D - 3)} {R_a}^c {R_c}^b {R_b}^a  + \frac{3D}{8(2D - 3)} R^3  \Bigg]\, ,
\end{align}
while ${\cal Z}_{(4)}$ and ${\cal Z}_{(5)}$ are presented in the appendix. Note these expressions are normalized in a way such that \req{sifj} holds.
The first two densities are proportional to the Einstein-Hilbert and Gauss-Bonnet terms. The  cubic-order one chosen here corresponds to the one in \cite{Quasi} up to a normalization. 

Interestingly, we find that a pattern analogous to the one found for the GQTGs in \req{recu1} holds for Quasi-topological gravities.
\begin{align}\label{zre}
{\cal Z}_{(n+5)} = -\frac{3(n+3){\cal Z}_{(1)} {\cal Z}_{(n+4)}}{D(D-1)(n+1)}  + \frac{3(n+4) {\cal Z}_{(2)}{\cal Z}_{(n+3)} }{D(D-1)n}- \frac{(n+3)(n+4) {\cal Z}_{(3)} {\cal Z}_{(n+2)}}{D(D-1)n(n+1)}\, .
\end{align}
Again, it is a computationally straightforward task to verify that replacing the explicit expressions for $\mathcal{Z}_{(m)}|_f$ with $m=1,2,3,(n+2),(n+3),(n+4)$ appearing in \req{sifj} in the rhs of \req{zre}, the resulting expression  agrees with the one corresponding to $\mathcal{Z}_{(n+5)}|_f$. Hence, starting with the first five densities, we can construct Quasi-topological gravities of arbitrarily high curvature orders. Observe that, in principle, the fact that \req{totaldQT}  and \req{zre} hold does not necessarily guarantee that the densities ${\cal Z}_{(n)}$ resulting from the recursive relations will be of the Quasi-topological class for general $n\geq 6$. However, it is not difficult to show that any density satisfying \req{sifj} automatically satisfies the Quasi-topological condition \req{qtg22}. Hence, all densities constructed using  \req{sifj} will indeed be of the Quasi-topological class.

Observe the striking similarity between \req{zre} and the analogous recursive relation in \req{recu1} corresponding to the GQTG densities. In fact, the only difference between both expressions is the ``4'' in the denominator of \req{recu1}, which appears replaced by a ``$D$'' in the analogous Quasi-topological expression. This does not seem to be a coincidence. When we go to four dimensions, both recursive relations match, and that is also the dimensionality for which Quasi-topological gravities cease to exist ---see below for more comments regarding this.

Let us finally mention that, to the best of our knowledge, no analogous recursive relation holds for the Lovelock densities.

\subsection{Completing the proof of \cite{Bueno:2019ltp}: all higher-curvature gravities as GQTGs}
In \cite{Bueno:2019ltp} it was shown that any gravitational effective action involving a general sum of higher-curvature invariants ---built from arbitrary contractions of the Riemann tensor and the metric--- can be transformed into a linear combination of GQTG (possibly including Quasi-topological) densities by a metric redefinition, provided at least one of such densities existed at every order in curvature ---a detailed proof of why this holds can be found in that paper. Our results in the last two subsections remove the ``provided...'' part of the result and complete its proof, which can therefore be formulated as  a theorem.

\begin{theorem}\label{th1}
 Any higher-derivative gravity Lagrangian involving an arbitrary sum of invariants constructed from the Riemann tensor and the metric can be mapped, order by order, to a sum of GQTG terms through a metric redefinition $g_{ab}=\tilde{g}_{ab}+\tilde{Q}_{ab}$, where $\tilde{Q}_{ab}$ is a symmetric tensor constructed from $\tilde{g}_{ab}$ and its derivatives. 
\end{theorem}

Therefore, given some higher-curvature effective action, it is always possible to map it to a GQTG frame in which the study of static black hole solutions is dramatically simplified. The equations which characterize the black hole solutions of general GQTGs (and Quasi-topological gravities) are fully  worked out in Section \ref{feqss}.

Note finally that, in fact, the results in \cite{Bueno:2019ltp} also hold (at least) for general densities involving eight (or less) derivatives of the metric (including covariant derivatives of the Riemann tensor), as well as for densities constructed from an arbitrary number of Riemann tensors and two covariant derivatives. Theorem \ref{th1} also holds when the corresponding gravitational effective action contains terms of that kind (and possibly for general densities involving arbitrary numbers of covariant derivatives and Riemann tensors).

\section{Explicit formulas for $n$-th order densities}\label{explifo}
In the previous section we provided recursive relations which allow for the construction of GQTG and Quasi-topological densities in general dimensions and for arbitrary curvature orders. However, those expressions may be somewhat messy to apply in practice if we want to construct very high order densities ---\eg a $n=1729$ Quasi-topological gravity. In this section we provide explicit expressions for $n$-th order densities of the GQTG and Quasi-topological classes. 

Our strategy here will be to construct general enough order-$n$ invariants built from general combinations of certain ``seed-invariants'' such that the coefficients can be chosen to satisfy the GQTG or Quasi-topological conditions respectively.
In order to do that, we find it  convenient to start from a set of tensors constructed as linear combinations of the form
\begin{equation}\label{KRRR}
\leftidx{^{(i)}}{\mathcal{K}}_{ab}^{cd}= \leftidx{^{(i)}}{c}_0 R_{ab}^{cd}- \leftidx{^{(i)}}{c}_1 R_{[a}^{[c}\delta_{b]}^{d]}- \leftidx{^{(i)}}{c}_2 R\delta_{[a}^{[c}\delta_{b]}^{d]}\, ,
\end{equation}
where $i=1,2,3$ and the constants $\leftidx{^{(i)}}{c}_{0,1,2}$ will be fixed in a moment. Before doing so, we observe that using \req{rieef} it follows that 
\begin{equation}
\left. \leftidx{^{(i)}}{\mathcal{K}}_{ab}^{cd} \right|_{f}=  \leftidx{^{(i)}}{K}_{1}T^{[a}_{[c}T^{b]}_{d]} +\leftidx{^{(i)}}{K}_{2} T^{[a}_{[c}\sigma^{b]}_{d]} + \leftidx{^{(i)}}{K}_{3}\sigma^{[a}_{[c}\sigma^{b]}_{d]}\, ,
\end{equation}
where
\begin{align}
{K}_{1}&\equiv (-2c_0 + c_1 + 2 c_2)A - (D-2)(c_1 + 4 c_2)B  -(D-2)(D-3) c_2 \psi \, , \\
{K}_{2}&\equiv (4 c_2 + c_1)A + (-D c_1 - 8(D-2) c_2 + 4 c_0) B - (D-3)(2 (D-2)c_2 + c_1) \psi  \, , \\
{K}_{3}&\equiv 2 c_2 A - 2 (2(D-2)c_2 +  c_1)B - ((D-3) c_1 - 2 c_0 + (D-2)(D-3)c_2) \psi\, , 
\end{align}
and where we omitted the $^{(i)}$ superindices everywhere to avoid some unnecessary clutter. This means that the invariants $\leftidx{^{(i)}}{\mathcal{K}}_{ab}^{cd}$ evaluated on the  metric (\ref{fEq}) can be written as a linear combination of products of projectors analogous to the one corresponding to the Riemann tensor itself. The relation between the  $\leftidx{^{(i)}}{c}_{0,1,2}$ and the  $\leftidx{^{(i)}}{K}_{1,2,3}$  allows us to move from the covariant expressions for the densities constructed from the  $\leftidx{^{(i)}}{\mathcal{K}}_{ab}^{cd}$, and their resulting expressions when evaluated on \req{fEq}.

Let us introduce some additional notation. We define
\begin{equation}\label{notaK}
\left(\leftidx{^{(i)}}{\mathcal{K}^q} \right)_{ab}^{cd}\equiv \leftidx{^{(i)}}{\mathcal{K}}_{ab}\,^{a_1 b_1}\, \leftidx{^{(i)}}{\mathcal{K}}_{a_1b_1}\,^{a_2 b_2} \dots \leftidx{^{(i)}}{\mathcal{K}}_{a_q b_q}\,^{cd}\, ,
\end{equation}
and, from these, the invariants
\begin{equation}\label{sseed}
\mathcal{R}_{q,m,p}\equiv  \left(\leftidx{^{(1)}}{\mathcal{K}^q} \right)_{ab}^{cd} \left(\leftidx{^{(2)}}{\mathcal{K}^m} \right)_{cd}^{ef}\left(\leftidx{^{(3)}}{\mathcal{K}^p} \right)_{ef}^{ab}\, .
\end{equation}
After some algebra, it is possible to show that when we evaluate these on 
the  metric (\ref{fEq}), they can be written as
\begin{align}
\left.{\cal R}_{q,m,p}\right|_f = &+\left(\leftidx{^{(1)}}{K_1}\right)^q\,\, \left(\leftidx{^{(2)}}{K_1}\right)^m \,\, \left(\leftidx{^{(3)}}{K_1}\right)^p \\ \notag &+\left(\leftidx{^{(1)}}{K_2}\right)^q\,\, \left(\leftidx{^{(2)}}{K_2}\right)^m \,\, \left(\leftidx{^{(3)}}{K_2}\right)^p  2^{1-(q+m+p)} (D-2)\\ \notag  &+ \left(\leftidx{^{(1)}}{K_3}\right)^q\,\, \left(\leftidx{^{(2)}}{K_3}\right)^m \,\, \left(\leftidx{^{(3)}}{K_3}\right)^p \frac{(D-2)(D-3)}{2} \, .
\end{align}
Next, we should choose the constants $\leftidx{^{(i)}}{c}_{0,1,2}$. We find it useful to fix them so that the resulting tensors possess the following factorization property: when evaluated on the  metric (\ref{fEq}), we would like them to be such that the dependence on the radial coordinate $r$ completely factorizes from the tensorial dependence, \ie they should be such that $\left. \leftidx{^{(i)}}{\mathcal{K}}_{ab}^{cd} \right|_f=\leftidx{^{(i)}}{F(r)} A_{ab}^{cd} $, where $A_{ab}^{cd} $ is independent from $r$. Using this criterion, we set
\begin{alignat}{4}
\leftidx{^{(1)}}{c_0} & \label{1c0} = 0 \,, \quad &&\leftidx{^{(1)}}{c_1} = -1 \, ,\quad &&&\leftidx{^{(1)}}{c_2}  = \frac{1}{D} \, ,  \\
\leftidx{^{(2)}}{c_0} & \label{2c0}  = 0 \,, \quad &&\leftidx{^{(2)}}{c_1} = 0 \, ,\quad &&&\leftidx{^{(2)}}{c_2}= -1 \, ,\\ 
\leftidx{^{(3)}}{c_0} & \label{3c0} = 1 \,, \quad &&\leftidx{^{(3)}}{c_1}= \frac{4}{(D-2)} \, ,\quad &&&\leftidx{^{(3)}}{c_2} =  \frac{2}{(D-1)(D-2)} \, .
\end{alignat}
%\begin{alignat}{4}
%\leftidx{^{(1)}}{c_0} & \label{1c0} = 1 \,, \quad &&\leftidx{^{(1)}}{c_1} = \frac{4}{(D-2)} \, ,\quad &&&\leftidx{^{(1)}}{c_2}  = -\frac{2}{(D-1)(D-2)} \, ,  \\
%\leftidx{^{(2)}}{c_0} & \label{2c0}  = 0 \,, \quad &&\leftidx{^{(2)}}{c_1} = -1 \, ,\quad &&&\leftidx{^{(2)}}{c_2}= \frac{1}{D} \, ,\\ 
%\leftidx{^{(3)}}{c_0} & \label{3c0} = 0 \,, \quad &&\leftidx{^{(3)}}{c_1}= 0 \, ,\quad &&&\leftidx{^{(3)}}{c_2} = -1 \, .
%\end{alignat}
With these choices, $\leftidx{^{(1)}}{\mathcal{K}}_{ab}^{cd}$, $\leftidx{^{(2)}}{\mathcal{K}}_{ab}^{cd}$ and $\leftidx{^{(3)}}{\mathcal{K}}_{ab}^{cd}$ are respectively the traceless Ricci tensor $R_a^b - (1/D)\delta_a^b R$ with each of its mixed indices antisymmetrized with those of a delta, the Ricci scalar times two deltas doubly antisymmetrized, and the Weyl tensor. Explicitly, when evaluated on \req{fEq} they read
\begin{equation}
\left.{\cal R}_{q,m,p}\right|_f = \left[A-(D-4)B + (D-3)\psi \right]^q \left[-2 A + (D-2)(4B + (D-3)\psi) \right]^m  [A + 2 B - \psi]^p  \beta_{p, q} \, ,
\end{equation}
where
\begin{align}
\beta_{p, q} \equiv& \frac{(-1)^q 2^p}{D^q \left[(D-1)(D-2) \right]^p} \left[(-1)^{p+q} 2^{p+q-1} (D-2)(D-3) + 2^{1-q} (D-2)(D-3)^p(D-4)^q \right.
\\ \notag
& \left. +  (-1)^p (D-2)^{p+q} (D-3)^p  \right] \, .
\end{align}
%\begin{align}
%\beta_{p,q}  \equiv  &+ 2^{1-q} (D-2)(D-3)^p(D-4)^q +(-1)^{p+q} 2^{p+q-1} (D-2)(D-3)  \\ \notag &+ (-1)^p  (D-2)^{p+q} (D-3)^p \, .
%\end{align}
Note that they identically vanish both in $D=2$ and $D=3$, with some degree of simplification also taking place for $D=4$. 

The $\mathcal{R}_{q,m,p}$ with the $\leftidx{^{(i)}}{c}_{0,1,2}$ chosen in this way will be the seed invariants we will use to construct explicit order-$n$ GQTG and Quasi-topological densities. The next step is then to consider a generic linear combination of the form
\begin{equation}\label{ks}
\mathcal{L}_{(n)}=\sum_{i=0}^n \sum_{j=0}^{n-i} \alpha_{i,j}^{(n)}{\cal R}_{j,n-i-j,i}\, ,
\end{equation}
the goal being to fix the $\alpha_{i,j}^{(n)}$ so that $\mathcal{L}_{(n)}$ satisfies the GQTG or Quasi-topological conditions for general $n$. Let us start with the former.

\subsection{Generalized quasi-topological gravities}
GQTG densities have been previously constructed for: $n=3$ and general $D\geq 4$ in \cite{Hennigar:2017ego}; $n=4$ and general $D\geq 4$ in \cite{Ahmed:2017jod}; $n=5,\cdots,10$ and $D=4$ in \cite{PabloPablo4,Arciniega:2018tnn}. Here we extend those results to general $n$ and $D\geq 4$.

In order for $\mathcal{L}_{(n)}$ to be of the GQTG class, we need to impose condition (\ref{totit}). A careful analysis reveals that we can reduce the number of terms appearing in the sum over $j$ in \req{ks} to only two. In particular, an expression of the form 
\be \label{snn}
{\cal S}_{(n)} = \sum_{i=0}^{n}  \frac{\alpha^{(n)}_{i, 0}}{\beta_{i,0}}{\cal R}_{0, n-i, i} +\sum_{i=0}^{n-2}  \frac{\alpha^{(n)}_{i, 2}}{\beta_{i,2}}{\cal R}_{2, n-i-2, i}  \, ,
\ee
suffices. We find that for
%\begin{align}
% \alpha^{(n)}_{i, 0} &= - \frac{2^{2(j-n) + i + 2}  (i-1)((i-1)D + 2 - i) n!}{(i+j)! (n-i-j)! (D-2) (D-1)^{n-j-1}} \, , \\
%  \alpha^{(n)}_{i, 2} &=   \frac{2^{2(j-n) + i}  (i+1)(i+2) n!}{(i+j)! (n-i-j)! (D-2) (D-1)^{n-j}}\, ,
%\end{align}
\begin{align}
\alpha^{(n)}_{i,0} &= - \frac{2^{2(1-n) + i} (i-1) [2 - i + D(i-1)] n!}{i!(n-i)! (D-1)^{n-1} (D-2)} \, ,
\\
\alpha^{(n)}_{i,2} &= \frac{2^{2(2-n) + i} (i+1)(i+2) n!}{(i+2)!(n - i - 2)! (D-1)^{n-2} (D-2)}\, ,
\end{align}
%\begin{align}
 %\alpha^{(n)}_{i, 0} &= -\frac{2^{2(1-n)+i} (-1)^{n+i} (i+1)  \left((i-1)D - (i-2)\right) n!}{ (D-1)^{(n-1)} (D-2) (n-i)! i!}\, , \\
  %\alpha^{(n)}_{i, 2} &= \frac{2^{2(2-n)} (-1)^{n+i}n!}{ (D-1)^{n-2} (D-2) (n-i-2)! i!} \, ,
%\end{align}
the GQTG condition is satisfied for general $n$. From the above we can see that the corresponding order-$n$ density will contain $2n-1$ terms (naively the sum involves $2n + 2$ terms, however three of those densities, corresponding to the combinations $(0,1)$, $(1,0)$ and $(1,1)$ vanish identically independent of the space-time dimension). Each one of those can be obtained from eqs. (\ref{sseed}), (\ref{notaK}) and \req{KRRR} with the $\leftidx{^{(i)}}{c}_{0,1,2}$ given in eqs. (\ref{1c0}), (\ref{2c0}) and (\ref{3c0}).

Defined as in \req{snn}, the ${\cal S}_{(n)} $ are GQTG densities normalized such that \req{densiexp} holds. While it is fair to say that the above expression is not particularly simple, it can be straightforwardly evaluated (at least in principle and perhaps with some computer help) at any order.

\subsection{Quasi-topological gravities}
Let us now move on to the case of Quasi-topological gravities. Recall that theories of this kind have been previously constructed for: $n=3$ and general $D\geq 5$ in \cite{Quasi,Quasi2}; $n=4$ and general $D\geq 5$ \cite{Dehghani:2013ldu,Ahmed:2017jod}; $n=5$ and $D=5$ in \cite{Cisterna:2017umf}. In addition to those, and prior to them, a particular subset of Quasi-topological gravities had been  identified for $D=2n-1$ and general $n$ in \cite{Oliva:2011xu} ---see \req{espQT} below. Just like for the GQTG ones, here we extend the construction of Quasi-topological gravities to general $n$ and $D\geq 5$.

Before doing so, let us recall the special Quasi-topological gravities for which explicit formulas are known. The first corresponds to Lovelock theories which, from our perspective, can be understood as a subset of the Quasi-topological class. Naturally, for those a closed-form expression is well-known to exist. This reads
\begin{equation}
\mathcal{X}_{(n)}=\frac{(2n)!}{2^n} \delta^{[a_1}_{b_1}\delta^{a_2}_{b_2}\cdots \delta_{b_{2n}}^{a_{2n}]}R_{a_1a_2}^{b_1b_2} \cdots R_{a_{2n-1}a_{2n}}^{b_{2n-1}b_{2n}}\, ,
\end{equation}
which reduces to the Euler density of compact manifolds when $D=2n$.

Also, a particular family of Quasi-topological gravities was constructed in $D=2n-1$ for arbitrary $n$ \cite{Quasi2,Oliva:2010zd,Oliva:2011xu}. In addition to condition (\ref{qtt}), theories belonging to this family are characterized by possessing second-order traced field equations. The corresponding densities are defined as
\begin{align}\label{espQT}
\tilde{\mathcal{Z}}_{(n)}=&\frac{(2n)!}{2^n (D-2n+1)} \delta^{[a_1}_{b_1}\delta^{a_2}_{b_2}\cdots \delta_{b_{2n}}^{a_{2n}]} \left(W_{a_1 a_2}^{b_1b_2}\cdots W_{a_{2n-1}a_{2n}}^{b_{2n-1}b_{2n}}-R_{a_1 a_2}^{b_1b_2}\cdots R_{a_{2n-1}a_{2n}}^{b_{2n-1}b_{2n}} \right)\\ \notag &- c_n W_{a_1 a_2}^{a_{2n-1}a_{2n}}W_{a_3 a_4}^{a_1 a_{2}}\dots W_{a_{2n-1} a_{2n}}^{a_{2n-3} a_{2n-2}}\, ,
\end{align}
where
\begin{equation}
c_n\equiv \frac{(D-4)!}{(D-2n+1)!}\frac{\left[n(n-2)D(D-3)+n(n+1)(D-3)+(D-2n)(D-2n-1) \right]}{\left[(D-3)^{n-1}(D-2)^{n-1}+2^{n-1}-2(3-D)^{n-1}\right]}\, .
\end{equation}

After reviewing these cases, let us now resume our construction. In the case of Quasi-topological gravities, in addition to \req{totit} we need to impose an additional condition, given by \req{qtg22}. This complicates a bit the problem. In particular, we seem to require all terms appearing in \req{ks}. Explicitly, we find
\be 
{\cal Z}_{(n)} = \sum_{i=0}^{n} \sum_{j=0}^{n-i} \frac{\alpha^{(n)}_{i, j}}{\beta_{i,j}} {\cal R}_{j, n-i-j, i} \, ,
\ee
where 
%\begin{align}
%\alpha^{(n)}_{i,j} \equiv&  \frac{(-1)^{n+1+j} 2^{2j-2} n!}{i! j! (n-i-j)!}\\ \notag &\times\frac{\left[4 i(i-1) - 4(i-1)(j+2i-1) D +  \left(2(i-1) + (j + 2i-3)(j+2i-2)  \right) D^2 \right]}{ D^{n-j} (D-1)^{n-j} (D-2)^{i+j}}\, .
%\end{align}
\begin{align}
\alpha^{(n)}_{i,j} \equiv&   \frac{(-1)^{i + 1} 2^{2j + i -2} n!}{i! j! (n-i-j)!}
\\\notag &\times \frac{\left[4 i(i-1) - 4(i-1)(j+2i-1) D +  \left(2(i-1) + (j + 2i-3)(j+2i-2)  \right) D^2 \right]}{ D^{n-i} (D-1)^{n-j} (D-2)^{i+j}}\, .
\end{align}
Defined like this, ${\cal Z}_{(n)}$ is a Quasi-topological density normalized so that \req{sifj} holds. Again, the expression is not particulary simple but, again, it is completely explicit and straightforward to evaluate at any order (at least in principle).

For the Quasi-topological gravities, our construction requires $(n-1)(n+4)/2$ densities (again the sum naively contains $(n+1)(n+2)/2$ densities, but the ones corresponding to the $(i, j)$ combinations $(1,0)$, $(0,1)$ and $(1,1)$ vanish in a dimension-independent manner).

Before closing let us note that for both the GQTG and Quasi-topological theories, our determination of the constants has been performed in a dimension independent fashion. This leaves open the possibility that for some orders $n$ and in some spacetime dimension the covariant expressions we have presented for GQTG or Quasi-topological theories could fail to exist. Such an occurrence would be reflected in the existence of ``dimension specific'' zeroes of the constants $\beta_{i,j}$. For example, specializing to $D = 4$, it follows that
\be 
\beta_{i,j}^{(D=4)} =  \frac{(-1)^{i+j} 2^{i + j}}{3^i} \left[1 + (-1)^j \right] \, .
\ee
Obviously this means that $\beta_{i,j} = 0$ for any odd value of $j$ in $D = 4$. While there is no issue in this case for GQTG theories (since in those cases we use only $j = 0$ and $j=2$), this precludes the existence of Quasi-topological theories of order $n \ge 3$ in $D = 4$.  Proving the absence of zeroes in higher dimensions is more challenging, though it can indeed be done ---we relegate this proof to Appendix~\ref{betaRoots}. Therefore, the expressions presented above correspond to genuine, covariant Quasi-topological Lagrangians for all $D > 4$ and all $n > 2$.

A further worry that one may have is that the expressions provided may become proportional to a Lovelock density (for all backgrounds) of some order and in some dimension. However, this cannot occur for the following reason. The expressions for the Quasi-topological Lagrangian densities presented here make use of only a single Weyl scalar of highest-order $n$. However, the Lovelock Lagrangian of order $n$ makes use of multiple independent Weyl scalars of order $n$, provided that $n > 2$ (as can be seen by re-writing the Lovelock Lagrangian in terms of the Weyl tensor). Since the Weyl scalars will differ on generic backgrounds, the Lagrangian densities cannot be equal.

\section{Black hole equations for $n$-th order densities}\label{feqss}
In Section \ref{GQTGss} we explained that in order to obtain the equation satisfied by the metric function $f(r)$ for a given density from the on-shell action method, we require the evaluation of $L_{N,f}$, as defined in \req{ansS}. In this section we show that there is an alternative route which only requires the use of the $g_{tt}g_{rr}=-1$ metric in \req{fEq}, and which allows us to write in full generality such equation for general GQTG and Quasi-topological densities of arbitrary order and for general $D$.

The idea is to make direct use of the full non-linear equations of motion, which we wrote above for any higher-curvature gravity in \req{eomGHOG}. Naturally, whenever the $g_{tt}g_{rr}=-1$ ansatz holds, such equations determine $f(r)$ when we evaluate them on it. An important piece of information is the expression for $\left.\tensor{P}{_{cd}^{ab}}\right|_f$ for a given density ${\cal R}_{(n)}$. After some algebra, it can be shown that this satisfies\footnote{Note that $P_{abcd}$ must have the form in \req{Pabcd} due to symmetry. The value of the coefficients $P_i$ can then be obtained using the identity $\delta {\cal L} = \tensor{P}{_{cd}^{ab}} \delta \tensor{R}{_{ab}^{cd}}$ applied to the case of the single-function metric \req{fEq}.}
\be \label{Pabcd}
\left.\tensor{P}{_{cd}^{ab}}\right|_{f} = P_1 T^{[a}_{[c} T^{b]}_{d]} +P_2   T^{[a}_{[c} \sigma^{b]}_{d]} +P_3 \sigma^{[a}_{[c} \sigma^{b]}_{d]} \, ,
\ee
where we defined
\begin{equation}
P_1\equiv  -  \frac{\partial {\cal R}_{(n)}|_f}{\partial f''} \, ,\quad P_2 \equiv - \frac{r}{D-2} \frac{\partial {\cal R}_{(n)}|_f}{\partial f'}  \, , \quad P_3\equiv - \frac{ r^2 }{(D-2)(D-3)} \frac{\partial {\cal R}_{(n)}|_f}{\partial f }\, .
\end{equation}
From this expression, we can work out that
\begin{align} 
\left.R_{ac}{}^{de} P_{de}{}^{bc}\right|_f &= \left[A  \frac{\partial  {\cal R}_{(n)}|_f}{\partial f''} - \frac{r B}{2}  \frac{\partial  {\cal R}_{(n)}|_f}{\partial f'}  \right] T_a^b + \left[\frac{r^2 \psi}{(D-2)}   \frac{\partial  {\cal R}_{(n)}|_f}{\partial f}-\frac{r  B}{(D-2)}  \frac{\partial  {\cal R}_{(n)}|_f}{\partial f'}   \right] \sigma_a^b\, .
\end{align}
We also need the expression resulting from taking two covariant derivatives of $P_{cd}{}^{ab}$. This results in\footnote{In deriving this expression we have already assumed that $\delta S_f/\delta f = 0$ holds and used this condition to simplify the result.}
\begin{align}\nonumber
\left.\nabla^c \nabla^d P_{acd}{}^b \right|_f= &-\frac{1}{8 r^2}\left[ r f' \left(2 r\frac{d P_1}{dr} + (D-2) (2P_1 -  P_2) \right)\right. \\ \nonumber & \quad\quad\quad\,  \left. + 2(D-2)f \left(r \frac{d P_2}{dr} + (D-3) (P_2 - 2 P_3) \right)\right]T_a^b 
\\ \nonumber
&- \frac{1}{ 4 r^2}\left[r^2 f \frac{d^2 P_2}{dr^2} + r f'\left( r \frac{d P_2}{dr} + (D-3) \left( P_2 - 2 P_3 \right) \right) \right. 
\nonumber\\ \label{ddP}
& \quad\quad\quad\, \left. + 2 (D-2) f \left( r \frac{d P_2}{dr} -  r \frac{dP_3}{dr} + \frac{(D-4)}{2} (P_2 - 2 P_3) \right)\right] \sigma_a^b\, .
\end{align}
With these expressions at hand, it is a matter of considering explicit expressions for the  $ {\cal R}_{(n)}$ in order to obtain the corresponding equations for $f(r)$ using \req{eomGHOG}.

\subsection{Generalized quasi-topological gravities}
Let us start with the GQTG densities. For those, a general-$n$ expression for the equation of $f(r)$ was presented for $D=4$ theories in \cite{PabloPablo4}. In that paper, such expression was guessed from its explicit computation for  $n=3,4,5$ and then verified to hold for $n=6,\dots,10$. In higher-dimensions, the equations are known for $n=3,4$ \cite{Hennigar:2017ego,Ahmed:2017jod}.

Using (\ref{eomGHOG}), (\ref{Pabcd}), (\ref{ddP}) and our explicit formula for $\mathcal{S}_{(n)}|_{f}$ in \req{densiexp}, we find that the equation of $f(r)$ can be integrated once, and we are left with \req{eqqqf}, where
\begin{align}\notag
\mathcal{F}_{{\cal S}_{(n)}} = \frac{1}{32}& \left(  \frac{-2 r f' + (D-4) (k-f)}{4 r^2} \right)^{n-3}\\ \notag & \times \bigg[ 4 n(n-1)(n-2)r^{D-5} f \big(r f' + 2(k-f) \big) f''   - 4 (n-1)(n-2) r^{D-4} f'^3
\nonumber\\ \notag
&+ 2 (n-2) r^{D-5} \Big(  \big((D-6)n^2 - 2 (D-7) n + 3(D-4)\big) f \\ \notag &+ k \big( (D-8)n - 3(D-4) \big) \Big) f'^2 
\nonumber\\ \notag
&+  \Big(\big(4(D-5) n^3 - 2 (7 D - 34) n^2 - (D^2 - 22 D + 80)n + 3(D-4)^2 \big)f \\ &+ k (n-3)(D-4)(D + 2n - 4) \Big)2 r^{D-6} (k-f) f'
\nonumber\\ \notag
&+ (D-4)r^{D-7} \Big( \big(8 n^3 - 24 n^2 - 2(D-12) n - (D-4)^2 \big) f  \\ \label{bhfG}
& +k (D-4)(D +2n -4) \Big) (k-f)^2 \bigg]\, .
\end{align}
%So the equation of $f(r)$ reads $F_{{\cal S}_{(n)}}=C$ for some constant $C$. 
This expression reduces to all previously known results in the corresponding particular cases, as it must.

Observe that the results in \cite{Bueno:2019ltp} along with Theorem \ref{th1} above make this equation particularly important in the following sense. Since we can map any gravitational effective action to a GQTG by a metric redefinition, the thermodynamic properties of black holes ---which are frame independent \cite{Jacobson:1993vj,Bueno:2019ltp}--- can be dramatically more easily studied in the GQTG frame. In such a frame, all information about the most general black hole is contained in \req{bhfG}. This would allow for a general study of the thermodynamics of higher-curvature gravity black holes in general $D$ along the lines of the four-dimensional case studied in \cite{PabloPablo4}.

\subsection{Quasi-topological gravities}
The case of Quasi-topological gravities is easier. This is because \req{qtg22} holds for them, and the field equations simply reduce to
\be 
\left.{\cal E}_a^b\right|_{f} = \left. R_{ac}{}^{de} P_{de}{}^{bc} - \frac{1}{2} \delta_a^b {\cal Z}_{(n)} \right|_{f}\, .
\ee
%One finds
%\begin{align} 
%\left.R_{ac}{}^{de} P_{de}{}^{bc}\right|_f &= \left[A  \frac{\partial  {\cal Z}_{(n)}|_f}{\partial f''} - \frac{r B}{2}  \frac{\partial  {\cal Z}_{(n)}|_f}{\partial f'}  \right] T_a^b + \left[\frac{r^2 \psi}{(D-2)}   \frac{\partial  {\cal Z}_{(n)}|_f}{\partial f}-\frac{r  B}{(D-2)}  \frac{\partial  {\cal Z}_{(n)}|_f}{\partial f'}   \right] \sigma_a^b\, .
%\end{align}
%This then suggests the following tentative form for the $r-r$ component of the field equation:
%\be 
%{\cal E}_r^r =  \frac{f'}{4 } \frac{\partial \cL}{\partial f'} + \frac{f''}{2}  %\frac{\partial \cL}{\partial f''} - \frac{\cL}{2} \, . 
%\ee
%This expression yields the correct field equations. For the particular lagrangians used above, the result is:
Using \req{sifj} for $ {\cal Z}_{(n)}|_f$, we are left with
\be 
\mathcal{F}_{{\cal Z}_{(n)}} = \frac{(D-2n)}{2} r^{D-2n-1} \left(k - f \right)^n \, ,
\ee
which is in agreement with the previously known expressions for $n=3,4,5$ in various dimensions \cite{Myers:2010jv,Dehghani:2013ldu,Ahmed:2017jod,Cisterna:2017umf}.  Note that while the above shows that ${\cal Z}_{(n)}$ does not correct the metric function in $D = 2 n$, the non-vanishing of the on-shell Lagrangian itself means that these terms can still prove non-trivial in some instances, \eg in the thermodynamic description of black holes.  This is similar to how Lovelock theory of order $n$ is dynamically trivial in $D = 2n$, yet still provides a non-trivial correction to the black hole entropy.

\subsection{Comments on densities normalization}

Up to this point we have normalized the Lagrangian densities  for the theories (equivalently, chosen the coupling constants) in a manner that makes the on-shell Lagrangian relatively simple to work with. However, this normalization is not canonical and indeed not the simplest normalization for working directly with the field equations. Therefore, before closing, we present some remarks on this point.

When the equations of motion for a given theory are evaluated on a maximally symmetric space they will always reduce to a polynomial equation
\be 
h(x) \equiv -\frac{2 L^2 \Lambda}{(D-1)(D-2)} - x + \sum_n c_n \lambda_n x^n = 0\, ,
\ee
that determines the radius of curvature of the maximally symmetric space in terms of the coupling constants of the given theory. Here $c_n$ are some constants and $x$ is related to the constant curvature of the maximally symmetric space, $R_{ab}^{cd} = -2 (x/L^2) \delta_{[a}^{[c} \delta_{b]}^{d]}$.  As discussed at length above, for Lovelock and Quasi-topological theories, this property of the field equations extends also to spherically symmetric spacetimes. Therefore a particularly convenient and simple normalization is that for which the constants $c_n$ are all equal to unity. For the Quasi-topological theories, this choice of normalization can be accomplished by the following rescaling:
\be 
{\cal Z}_{(n)} \to  \frac{(-1)^{n+1} (D-2) L^{2n -2} }{D-2n}{\cal Z}_{(n)} \, .
\ee
Note that this rescaling is singular in the dimension $D = 2n$. That is simply because Quasi-topological theories of order $n$ do not contribute to the equations of motion for maximally (and spherically) symmetric spacetimes in $D = 2n$. 

In the case of GQTGs, the equations of motion for the spherically symmetric spacetimes are no longer polynomial. However, it can still prove convenient to normalize these theories in the same manner described just above. For these theories, this amounts to the rescaling
\be 
{\cal S}_{(n)} \to \frac{(-1)^{n+1} 4^{n-1}(D-2) L^{2n-2}}{D^{n-1} (D-2n)} {\cal S}_{(n)} \, .
\ee 
Again, the factor of $D - 2n$ appears in the denominator as expected since the field equations for densities of order $n$ do not contribute to the equations of motion for maximally symmetric spaces in $D = 2n$. However, in this case it should be noted GQTG theories of order $n$ \textit{do} contribute non-trivially to the equations of motion of spherically symmetric spacetimes in dimension $D = 2n$. Therefore, the above normalization while always fine in $D = 4$, should not be applied in $D = 2n$ for $D \ge 4$. 

In dimensions larger than $4$ one has both Quasi-topological and GQTGs. Therefore, an alternative normalization for the GQTG theories would be such that they simply do not contribute to the equations determining the curvature scale of a maximally symmetric space. This normalization can be accomplished by subtracting from a particular GQTG density a term proportional to a Quasi-topological density of the same order:
\be 
{\cal S}_{(n)} \to {\cal S}_{(n)} - \left(\frac{D}{4}\right)^{n-1} {\cal Z}_{(n)} \, .
\ee

Finally, it is worth mentioning a final ambiguity in the choice of the coupling constants. The (Generalized) Quasi-topological theories are characterized by their properties on spherically symmetric spacetimes. However, this then means that the properties of these theories are left unchanged if one adds to the action a term that is trivial on these geometries. However, while this ambiguity has no effects in the realm of spherical symmetry (modulo a potential rescaling of the coupling), it could have important consequences for other geometries of interest, \eg in the cosmological context~\cite{Arciniega:2018fxj, Arciniega:2018tnn}.

\section{Final comments} \label{discuss}
 In this paper we have uncovered the structure of general-dimension and general-order GQTG and Quasi-topological gravities. More explicitly, we have shown that GQTG and Quasi-topological densities exist at all orders and how to obtain explicit expressions for such densities, as well as recursive relations between different-order densities. We have also obtained the general equations satisfied by the metric function $f(r)$ for general $D$ and $n$ in both cases. 
 
 Our results complete the proof presented in \cite{Bueno:2019ltp} that any higher-curvature effective action can be mapped via metric redefinitions to a GQTG. In particular, the thermodynamic properties of black holes for such general effective actions can be in principle obtained from their much simpler GQTG counterparts. The equations determining such solutions for general $D$ and $n$ are precisely the ones obtained in Section \ref{feqss} here, so it would be interesting to perform a full study of the thermodynamic properties of the black hole solutions of general $D$ and $n$  GQTG and Quasi-topological gravities, along the lines of the one performed in \cite{PabloPablo4} for the $D=4$ case. It would also be interesting to study the holographic dictionary for those classes of theories, along the lines of \cite{Buchel:2009sk,Camanho:2009vw,deBoer:2009pn,Myers:2010jv,Camanho:2010ru,Bueno:2018xqc,Mir:2019rik,Mir:2019ecg}. 
 
 By now, we know that special types of GQTGs satisfy additional properties, such as possessing second-order equations also for Taub-NUT/Bolt \cite{Bueno:2018uoy} and FLRW \cite{Arciniega:2018fxj,Cisterna:2018tgx,Arciniega:2018tnn} backgrounds. It would be interesting to verify whether the recursive relations constructed here give rise to higher-order densities satisfying these conditions when one starts with lower-order densities which do. Also, one could try to construct general expressions for densities satisfying these additional properties. Determining whether the order-reduction phenomenon occurring  in the equations extends to other kinds of solutions would also be an interesting problem.  Work along some of these lines is in progress.
 
% Let us close the paper with some prospects for future research.
 
  %\comment{the list below is really telling people a lot of possible projects; we should make sure we want to tell all this so explicitly }
  
%On the one hand, for the reasons explained in the paragraph above, it would be interesting to perform a full study of the thermodynamic properties of the black hole solutions of general $D$ and $n$  GQTG and Quasi-topological gravities, along the lines of the one performed in \cite{PabloPablo4} for the $D=4$ case. 

% By now, we know that special types of GQTGs satisfy additional properties, such as possessing second-order equations also for Taub-NUT/Bolt \cite{Bueno:2018uoy} and FLRW \cite{Arciniega:2018fxj,Cisterna:2018tgx,Arciniega:2018tnn} backgrounds. It would be interesting to verify whether the recursive relations constructed here give rise to higher-order densities satisfying these conditions when one starts with lower-order densities which do. Also, one could try to construct general expressions for densities satisfying these additional properties.
 
% In that regard, it could be that the order-reduction phenomenon occurring  in the equations extends to other kinds of solutions, such as rotating black holes.

%Finally, we suspect that a Birkhoff theorem should hold for general Quasi-topological gravities of arbitrary order, just like it does in the cases considered so far \cite{}. It should be possible to verify this statement using the results presented here.  

\section*{Acknowledgments}
We thank Graham Cox, Michael Deveau, Julio Oliva and Robb Mann for useful comments and suggestions. RAH thanks Mengqi Lu for previous collaboration on related topics.  In some cases, calculations performed in the manuscript have been facilitated by Maple and Mathematica, utilizing the specialized packages GRTensor and xAct~\cite{xact}. PB and RAH are grateful to the Yukawa Institute for Theoretical Physics for hospitality during the early stages of this project. The work of PB was supported by the Simons foundation through the It From Qubit Simons collaboration. The work of PAC is funded by Fundaci\'on la Caixa through a ``la Caixa - Severo Ochoa'' International pre-doctoral grant. PAC was further supported by the MCIU, AEI, FEDER (UE) grant PGC2018-095205-B-I00 and by the Spanish Research Agency (Agencia Estatal de Investigaci\'on) through the grant IFT Centro de Excelencia Severo Ochoa
SEV-2016-0597.  The work of RAH is supported physically by planet Earth through the electromagnetic and gravitational interactions, and by the Natural Sciences and Engineering Research Council of Canada through the Banting Postdoctoral Fellowship programme. 
 
\appendix

\section{Explicit Lagrangian densities for $n=4$ and $n=5$}
In this appendix we present the explicit form of the quartic and quintic GQTG and Quasi-topological densities used in Section \ref{recrel} to construct the recursive relations in eqs. (\ref{recu1}) and (\ref{z1}).
%\comment{I have independently verified that the expressions below produce the correct on-shell Lagrangians and field equations.  The $n=5$ cases have been printed directly from Mathematica using the `TexPrint' function to eliminate the possibility for typographical errors.}

\subsection{Generalized quasi-topological gravities}
{ \small
\begin{align}
{\cal S}_{(4)} =& +\frac{3 D}{8 (D-1)^2 (D-2)^2} R^2 R_{ab}R^{ab} \nonumber\\
&- \frac{(D^2 + 20 D - 20)}{64 (D-1)^3(D-2)^2} R^4 - \frac{3}{2(D-1)(D-2)(D-3)} R R^{ab}R^{cd}W_{acbd} 
\nonumber\\
&- \frac{3(2D^5 - 17 D^4 + 33 D^3 + 16 D^2 - 64 D + 32)}{32(D-1)^2(D-2)(D-3)(2D^4 - 17D^3 + 49 D^2 - 48 D + 16)} R^2 W_{abcd}W^{abcd} 
\nonumber\\
&- \frac{3 D}{(D-3) (2 D^4 - 17 D^3 + 49 D^2 - 48 D + 16)} R R^{ab} W_a{}^{cde}W_{bcde} 
\nonumber\\
&+ \frac{3 D^2}{4(D-3)(2 D^4 - 17 D^3 + 49 D^2 - 48 D + 16)} R^{ab}R^{cd}W_{ac}{}^{ef}W_{bdef} 
\nonumber\\
&+ \frac{(D-2)(2D-1)}{8(D-1)(D-3)(D^3 - 9 D^2 + 26 D - 22)} R W_{ab}{}^{ef}W^{abcd} W_{cdef} 
\nonumber\\
&+ \frac{3 D^2}{4 (D-3)(2 D^4 - 17 D^3 + 49 D^2 - 48 D + 16)} R_a^c R^{ab}W_{b}{}^{def}W_{cdef} 
\nonumber\\
&- \frac{3(D-2)^2(3D-2)}{64(D-3)(D^5 - 14 D^4 + 79 D^3 - 224 D^2 + 316 D - 170)} W_{ab}{}^{ef} W_{cd}{}^{ab}W_{gh}{}^{cd}W_{ef}^{gh}\, ,
\\
{\cal S}_{(5)} =&+ \frac{5 D R_{ab} R^{ab} R^3}{32 (-2 + D)^2 (-1 + D)^3} -  \frac{(-36 + 36 D + D^2) R^5}{256 (-2 + D)^2 (-1 + D)^4} -  \frac{15 R^2 R^{ab} R^{cd} W_{acbd}}{16 (-3 + D) (-2 + D) (-1 + D)^2} 
\nonumber\\
&-  \frac{5 (96 - 224 D + 144 D^2 + D^3 - 17 D^4 + 2 D^5) R^3 W_{abcd} W^{abcd}}{128 (-3 + D) (-2 + D) (-1 + D)^3 (16 - 48 D + 49 D^2 - 17 D^3 + 2 D^4)} 
\nonumber\\
&-  \frac{15 D R^{ab} R^2 W_{a}{}^{cde}W_{bcde}}{4 (-3 + D) (-1 + D) (16 - 48 D + 49 D^2 - \
17 D^3 + 2 D^4)} 
\nonumber\\
&+ \frac{15 D^2 R^{ab} R^{cd} R W_{ac}{}^{ef}W_{bdef}}{16 (-3 + D) (-1 + D) (16 - 48 D + 49 D^2 - 17 D^3 + 2 D^4)} 
\nonumber\\
&+ \frac{5 (-2 + D) D (392 - 1311 D + 1672 D^2 - 1037 
D^3 + 342 D^4 - 58 D^5 + 4 D^6) R^2 W_{ab}{}^{ef} W^{abcd} W_{cdef}}{64 (-3 + D) (-1 + D)^2 (-22 + 26 D - 9 D^2 + D^3) (176 - 600 D + 775 D^2 \
- 482 D^3 + 161 D^4 - 28 D^5 + 2 D^6)} 
\nonumber\\
&+ \frac{15 D^2 R_{a}{}^{c} R^{ab} R W_{b}{}^{def} W_{cdef}}{16 (-3 + D) (-1 + D) (16 - \
48 D + 49 D^2 - 17 D^3 + 2 D^4)} 
\nonumber\\
&+ \frac{5 (-2 + D) D R^{ab} R W_{a}{}^{cde} W_{bc}{}^{fg} W_{defg}}{4 (-3 + D) 
(176 - 600 D + 775 D^2 - 482 D^3 + 161 D^4 - 28 D^5 + 2 D^6)} 
\nonumber\\
&-  
\frac{5 (-2 + D) D^2 R^{ab} R^{cd} W_{ac}{}^{ef} W_{bd}{}^{gh} W_{efgh}}{16 (-3 + D) (176 - 600 D + 775 D^2 - 482 D^3 + 161 D^4 - 28 D^5 + 2 D^6)} 
\nonumber\\
&-  \frac{15 (-2 + D)^2 (-2 + 3 D) R W_{ab}{}^{ef} W^{abcd} 
W_{cd}{}^{gh} W_{efgh}}{256 (-3 + D) (-1 + D) (-170 + 316 D - 224 D^2 + 79 D^3 - 14 D^4 + D^5)} 
\nonumber\\
&-  
\frac{5 (-2 + D) D^2 R_{a}{}^{c} R^{ab} W_{b}{}^{def} W_{cd}{}^{gh} 
W_{efgh}}{16 (-3 + D) (176 - 600 D + 775 D^2 - 482 D^3 + 161 D^4 - 28 D^5 + 2 D^6)} 
\nonumber\\
&+ \frac{(-2 + D)^3 (-3 + 4 D) W_{ab}{}^{ef} W^{abcd} W_{cd}{}^{gh} W_{ef}{}^{if} W_{ghif}}{64 (-3 + D) (-1150 + 2954 D - 3202 D^2 + 1934 D^3 - 705 \
D^4 + 155 D^5 - 19 D^6 + D^7)}\, .
\end{align}
}
\subsection{Quasi-topological gravities}

{\small
\begin{align}
{\cal Z}_{(4)} =& -\frac{384 (D-8) R^a_b R_a^c R_c^d R^b_d}{(D-2)^5(D^3 - 8 D^2 + 48 D - 96)} - \frac{1152 R_{ab}R^{ab}R_{cd}R^{cd}}{(D-2)^5(D^3 - 8 D^2 + 48 D - 96)} 
\nonumber\\
&- \frac{64(D^3 - 10D^2 + 40D + 24) R R_a^c R_c^b R_b^a }{(D-1)(D-2)^5(D^3 - 8 D^2 + 48 D - 96)} + \frac{24(D^4 - 6 D^3 + 20 D^2 + 104 D - 64) R^2 R_{ab}R^{ab}}{(D-1)^2(D-2)^5(D^3 - 8 D^2 + 48 D - 96)} 
\nonumber\\
&- \frac{(D^5 + 6 D^4 - 64 D^3 + 416 D^2 + 176 D - 480) R^4}{(D-1)^3(D-2)^5(D^3 - 8 D^2 + 48 D - 96)} - \frac{96(D+2) R R^{ab}R^{cd} W_{acbd}}{(D-1)(D-2)^4(D-3)(D-4)} 
\nonumber\\
&- \frac{6(2D^5 - D^4 - 31 D^3 + 20 D^2 + 20 D - 16) R^2 W_{abcd}W^{abcd}}{(D-1)^2(D-2)^3(D-3)(D-4)(2D^4 - 17 D^3 + 49 D^2 - 48 D + 16)} 
\nonumber\\
&+ \frac{96(2 D^4 - 7 D^3 - 7 D^2 + 18 D - 8) R R_b^a W_{ac}{}^{de} W_{de}{}^{bc}}{(D-1)(D-2)^3(D-3)(D-4)(2 D^4 - 17 D^3 + 49 D^2 - 48 D + 16)} 
\nonumber\\
&+ \frac{384 R_a^c R^{ab} R^{de} W_{bdce}}{(D-2)^4(D-3)(D-4)} - \frac{48 (7D^2 - 10D + 4) R^{ab}R^{cd}W_{ac}{}^{ef} W_{bd ef}}{(D-2)^3(D-3)(2 D^4 - 17 D^3 + 49 D^2 - 48 D + 16)} 
\nonumber\\
&- \frac{8 (2 D^4 - 15 D^3 + 26 D^2 + 27 D - 58) R W_{ab}{}^{ef} W^{abcd} W_{cdef}}{(D-1)(D-2)^2(D-3)(D-4) (D^2 - 6 D + 11)(D^3 - 9 D^2 + 26 D - 22)} 
\nonumber\\
&- \frac{48 (7 D^2 - 10 D + 4) R_a^c R^{ab} W_b{}^{def}W_{cdef}}{(D-2)^3(D-3)(2 D^4 - 17 D^3 + 49 D^2 - 48 D + 16)} 
\nonumber\\
&+ \frac{96 R^{ab} W_a{}^{cde}W_{bc}{}^{fg}W_{defg}}{(D-2)^2(D-3)(D-4)(D^2 - 6 D + 11)} 
\nonumber\\
&- \frac{3(3D-4) W_{ab}{}^{cd} W_{cd}{}^{ef} W_{ef}{}^{gh} W_{gh}{}^{ab}}{(D-2)(D-3) (D^5 - 14 D^4 + 79 D^3 -224 D^2 + 316 D - 170)}\, ,
\end{align}
}
{ \tiny
\begin{align}
{\cal Z}_{(5)} =& +\frac{512 (-64 - 12 D + D^2) R_{a}{}^{c} R^{ab} R_{b}{}^{d} R_{c}{}^{e} R_{de}}{(-4 + D) (-2 + D)^6 (-128 + 32 D + D^3)} + \frac{5120 (4 + D) R_{ab} R^{ab} R_{c}{}^{e} R^{cd} R_{de}}{(-4 + D) (-2 + D)^6 (-128 + 32 D + D^3)} 
\nonumber\\
&-  \frac{640 (4608 + 3712 D - 2880 D^2 + 664 D^3 - 126 D^4 + 7 D^5) R_{a}{}^{c} R^{ab} R_{b}{}^{d} R_{cd} R}{(-4 + D) (-2 + D)^6 (-1 + D) (-128 + 32 D + D^3) (-96 + 48 D - 8 D^2 + D^3)}
\nonumber\\
& -  \frac{1920 (-768 - 320 D + 280 D^2 - 58 D^3 + 11 D^4) R_{ab} R^{ab} R_{cd} R^{cd} R}{(-4 + D) (-2 + D)^6 (-1 + D) (-128 + 32 D + D^3) (-96 + 48 D - 8 D^2 + D^3)} 
\nonumber\\
&-  \frac{160 (4096 - 27136 D + 7168 D^2 + 160 D^3 - 64 D^4 + 116 D^5 - 16 D^6 + D^7) R_{a}{}^{c} R^{ab} R_{bc} R^2}{(-4 + D) (-2 + D)^6 (-1 + D)^2 (-128 + 32 D + D^3) (-96 + 48 D - 8 D^2 + D^3)} 
\nonumber\\
&+ \frac{40 (30720 - 20992 D - 41216 D^2 + 17920 D^3 - 2784 D^4 + 656 D^5 + 28 D^6 - 8 D^7 + D^8) R_{ab} R^{ab} R^3}{(-4 + D) (-2 + D)^6 (-1 + D)^3 (-128 + 32 D + D^3) (-96 + 48 D - 8 D^2 + D^3)} 
\nonumber\\
&-  \frac{(155648 + 231424 D - 530176 D^2 + 136384 D^3 - 14336 D^4 + 4272 D^5 + 1296 D^6 - 204 D^7 + 16 D^8 + D^9) R^5}{(-4 + D) (-2 + D)^6 (-1 + D)^4 (-128 + 32 D + D^3) (-96 + 48 D - 8 D^2 + D^3)}
\nonumber\\
& -  \frac{240 (-80 - 100 D + 8 D^2 + 7 D^3) R^{ab} R^{cd} R^2 W_{acbd}}{(-4 + D) (-3 + D) (-2 + D)^5 (-1 + D)^2 (96 - 48 D + 7 D^2)} 
\nonumber\\
&-  \frac{10 (-128 + 896 D - 2552 D^2 + 3900 D^3 - 2970 D^4 + 425 D^5 + 710 D^6 - 243 D^7 - 10 D^8 + 8 D^9) R^3 W_{abcd} W^{abcd}}{(-4 + D) (-3 + D) (-2 + D)^4 (-1 + D)^3 (16 - 48 D + 49 D^2 - 17 D^3 + 2 D^4) (16 - 56 D + 69 D^2 - 27 D^3 + 4 D^4)} 
\nonumber\\
&+ \frac{240 (64 - 256 D + 260 D^2 + 292 D^3 - 795 D^4 + 516 D^5 - 35 D^6 - 42 D^7 + 8 D^8) R^{ab} R^2 W_{a}{}^{cde} W_{bcde}}{(-4 + D) (-3 + D) (-2 + D)^4 (-1 + D)^2 (16 - 48 D + 49 D^2 - 17 D^3 + 2 D^4) (16 - 56 D + 69 D^2 - 27 D^3 + 4 D^4)} 
\nonumber\\
&+ \frac{1920 (-48 - 14 D + 7 D^2) R_{a}{}^{c} R^{ab} R^{de} R W_{bdce}}{(-4 + D) (-3 + D) (-2 + D)^5 (-1 + D) (96 - 48 D + 7 D^2)} 
\nonumber\\
&-  \frac{240 (128 - 704 D + 1464 D^2 - 1240 D^3 + 60 D^4 + 503 D^5 - 221 D^6 + 28 D^7) R^{ab} R^{cd} R W_{ac}{}^{ef} W_{bdef}}{(-4 + D) (-3 + D) (-2 + D)^4 (-1 + D) (16 - 48 D + 49 D^2 - 17 D^3 + 2 D^4) (16 - 56 D + 69 D^2 - 27 D^3 + 4 D^4)}
\nonumber\\
& -  \frac{11520 R_{a}{}^{c} R^{ab} R_{d}{}^{f} R^{de} W_{becf}}{(-3 + D) (-2 + D)^5 (96 - 48 D + 7 D^2)} 
\nonumber\\
&-  \frac{20 (3632 - 7644 D - 4296 D^2 + 23905 D^3 - 23526 D^4 + 8466 D^5 + 560 D^6 - 1437 D^7 + 478 D^8 - 70 D^9 + 4 D^{10}) R^2 W_{ab}{}^{ef} W^{abcd} W_{cdef}}{(-4 + D) (-3 + D) (-2 + D)^3 (-1 + D)^2 (11 - 6 D + D^2) (-22 + 26 D - 9 D^2 + D^3) (176 - 600 D + 775 D^2 - 482 D^3 + 161 D^4 - 28 D^5 + 2 D^6)}
\nonumber\\
& -  \frac{240 (128 - 704 D + 1464 D^2 - 1240 D^3 + 60 D^4 + 503 D^5 - 221 D^6 + 28 D^7) R_{a}{}^{c} R^{ab} R W_{b}{}^{def} W_{cdef}}{(-4 + D) (-3 + D) (-2 + D)^4 (-1 + D) (16 - 48 D + 49 D^2 - 17 D^3 + 2 D^4) (16 - 56 D + 69 D^2 - 27 D^3 + 4 D^4)} 
\nonumber\\
&-  \frac{15360 R_{a}{}^{c} R^{ab} R_{b}{}^{d} R^{ef} W_{cedf}}{(-3 + D) (-2 + D)^5 (96 - 48 D + 7 D^2)} 
\nonumber\\
&+ \frac{960 (4 - 12 D + 11 D^2) R_{a}{}^{c} R^{ab} R^{de} W_{bd}{}^{fg} W_{cefg}}{(-4 + D) (-3 + D) (-2 + D)^4 (16 - 56 D + 69 D^2 - 27 D^3 + 4 D^4)} 
\nonumber\\
&+ \frac{160 (-232 + 550 D - 253 D^2 - 242 D^3 + 221 D^4 - 62 D^5 + 6 D^6) R^{ab} R W_{a}{}^{cde} W_{bc}{}^{fg} W_{defg}}{(-4 + D) (-3 + D) (-2 + D)^3 (-1 + D) (11 - 6 D + D^2) (176 - 600 D + 775 D^2 - 482 D^3 + 161 D^4 - 28 D^5 + 2 D^6)} 
\nonumber\\
&+ \frac{320 (4 - 12 D + 11 D^2) R_{a}{}^{c} R^{ab} R_{b}{}^{d} W_{c}{}^{efg} W_{defg}}{(-4 + D) (-3 + D) (-2 + D)^4 (16 - 56 D + 69 D^2 - 27 D^3 + 4 D^4)} 
\nonumber\\
&-  \frac{80 (12 - 28 D + 17 D^2) R^{ab} R^{cd} W_{ac}{}^{ef} W_{bd}{}^{gh} W_{efgh}}{(-3 + D) (-2 + D)^3 (176 - 600 D + 775 D^2 - 482 D^3 + 161 D^4 - 28 D^5 + 2 D^6)} 
\nonumber\\
&-  \frac{15 (-528 + 482 D + 241 D^2 - 425 D^3 + 194 D^4 - 39 D^5 + 3 D^6) R W_{ab}{}^{ef} W^{abcd} W_{cd}{}^{gh} W_{efgh}}{(-4 + D) (-3 + D) (-2 + D)^2 (-1 + D) (85 - 99 D + 48 D^2 - 11 D^3 + D^4) (-170 + 316 D - 224 D^2 + 79 D^3 - 14 D^4 + D^5)} 
\nonumber\\
&-  \frac{80 (12 - 28 D + 17 D^2) R_{a}{}^{c} R^{ab} W_{b}{}^{def} W_{cd}{}^{gh} W_{efgh}}{(-3 + D) (-2 + D)^3 (176 - 600 D + 775 D^2 - 482 D^3 + 161 D^4 - 28 D^5 + 2 D^6)} 
\nonumber\\
&+ \frac{240 R^{ab} W_{a}{}^{cde} W_{bc}{}^{fg} W_{de}{}^{hi} W_{fghi}}{(-4 + D) (-3 + D) (-2 + D)^2 (85 - 99 D + 48 D^2 - 11 D^3 + D^4)}
\nonumber\\
& -  \frac{4 (-5 + 4 D) W_{ab}{}^{ef} W^{abcd} W_{cd}{}^{gh} W_{ef}{}^{ij} W_{ghij}}{(-3 + D) (-2 + D) (-1150 + 2954 D - 3202 D^2 + 1934 D^3 - 705 D^4 + 155 D^5 - 19 D^6 + D^7)}\, .
\end{align}
}

\section{Behaviour of $\beta_{p,q}$ for $D > 4$}
\label{betaRoots}
Here we will show that $\beta_{p,q}$ cannot have any dimension specific zeroes for $D > 4$. By this we mean that for no values of $p,q$ does $\beta_{p,q}$ admit a factor $(D-N)$ for some integer $N >  4$, therefore indicating that the covariant expressions for quasi-topological Lagrangians presented in the main text are valid for arbitrary $n > 2$ in arbitrary $D > 4$.

Our argument will make use of Descartes' rule of signs. Recall that this rule indicates that when a polynomial is  ordered by descending variable coefficients the number of real, positive roots is either equal to the number of sign differences between consecutive non-zero coefficients, or differs from it by an even number. Certainly a necessary condition for $\beta_{p,q}$ to admit a dimension-specific zero would be for it to admit a zero for some real, positive $D > 4$. Therefore, our objective will be to prove that this cannot be the case. 

Written in terms of the quantity $x= D-4$, the factor in $\beta_{p,q}$ becomes
\be 
2 (x+1)^p \left[(-1)^p 2^q (x+2)^{p+q} + 4 x^q + 2 x^{q+1} \right] + (-1)^{p+q} 4^q 2^p \left[2 + 3 x + x^2 \right]\, ,
\ee
where we have included an extra factor of $2^{q+1}$ to ensure that all the coefficients are integers. We will now show that in fact the coefficients in this object always have sign governed by $(-1)^p$. First, by examining the positions of the $(-1)^p$ and $(-1)^{p+q}$ factors, we can recognize fairly easily the places where sign flips must occur if they do at all: Between the coefficients of $x^{q+2}$ and $x^{q+1}$, between $x^{q+1}$ and $x^q$, between $x^{q}$ and $x^{q-1}$, between $x^3$ and $x^2$, between $x^2$ and $x$, or finally between $x$ and the constant term. We must work through a few cases:
\begin{description}
\item[Case I:] $q > 2, p \ge 0$. For the terms $x^q$ and $x^{q+1}$ it is sufficient to look just at the terms in the square brackets and ignore the overall factor of $(1+x)^p$. In this case, the coefficient of $x^{q+1}$ in the square brackets is 
\be 
(-1)^p 2^{p + q -1} \frac{(p+q)!}{(q+1)! (p-1)!} (1 - \delta_{p,0}) + 2  \, ,
\ee
where the $\delta_{p,0}$ is the Kronecker symbol, indicating here that the first term is present only for $p > 0$. We see that the term never vanishes, and always has sign equal to that of $(-1)^p$.  The coefficient of $x^q$ is 
\be 
(-1)^p 2^{p + q} \frac{(p+q)!}{q! p!} + 4 \, ,
\ee
which again never vanishes and has sign equal to that of $(-1)^p$. Next observe that the coefficient of $x^2$ is:
\be
(-1)^p 2^{2q + p} \left[p(p-1) + \frac{(p+q)(p+q-1)}{4} + p(p+q)  + (-1)^q \right] \, .
\ee
Again, under the present assumptions this quantity does not vanish and always has sign of $(-1)^p$. The coefficient of $x$ is:
\be
(-1)^p 2^{2q + p} \left[3p + q + 3 (-1)^q  \right] \, ,
\ee
which can vanish for the case $q = 3$ if $p=0$, but otherwise has sign $(-1)^p$. Finally, the constant term is
\be 
2 (-1)^p 2^{2q + p} \left[1 + (-1)^q \right] \, ,
\ee
it vanishes for odd $q$, but when non-zero always has sign $(-1)^p$. This then proves that in the case $q > 2$ and $p \ge 0$ the entire object has sign $(-1)^p$. Therefore by Descartes' rule of signs there can be no positive roots, and therefore no dimension dependent zeros other than $D = 4$ under these assumptions.
\item[Case II:] $q = 2, p \ge 0$. In this case, we need to consider the coefficient of $x^3$ from inside the first set of square brackets. This coefficient is:
\be 
 \frac{(-1)^p 2^{p}(p+2)(p+1)p}{3}  + 2 \, .
\ee
This term always has sign $(-1)^p$ as when $p=0$ it is positive, and for $p \ge 1$, the first term is always larger in magnitude than the second and so the sum of the two has sign $(-1)^p$. The coefficient of $x^2$ is given by:
\be 
(-1)^p 2^{4 + p} \left[1 + p(p-1) + \frac{(p+1)(p+2)}{4} +  p(p+2) \right] + 8 \, .
\ee
Again, it is easy to see that the first term here always overwhelms the second, and the overall sign is $(-1)^p$. The coefficient of $x$ is
\be 
(-1)^p 2^{4 + p} \left[3 p + 5 \right]\, ,
\ee
which has sign $(-1)^p$. And finally, the constant term is given by
\be 
64 (2)^p (-1)^p \, ,
\ee
which obviously has the correct behaviour. Therefore, for the case $q = 2$ and $p \ge 0$, the coefficients of the polynomial are always of the sign $(-1)^p$ and there are no dimension-specific zeros for $D > 4$.

\item[Case III:] $q  = 1, p \ge 0$. Here, the coefficient of $x^2$ is
\be 
(-1)^p 2^{2 + p} \left[p(p-1) +  \frac{p(p+1)}{4} + p(p+1)  - 1  \right]  + 4(1+2p) \, . 
\ee
When $p = 0$, the term in brackets in negative, but the $+4$ ensures the coefficient is positive. When $p > 0 $, it is clear that the term in brackets dominates. Thus, the sign is always $(-1)^p$. The coefficient of $x$ is:
\be 
(-1)^p 2^{2 + p} \left[3p - 2  \right]  + 8 \, .
\ee
In this case, the $p=1$ term vanishes, but when non-vanishing the coefficient is always of sign $(-1)^p$. In this case the constant term vanishes. Therefore, for the case $q = 1$ and $p \ge 0$, the coefficients of the polynomial are always of the sign $(-1)^p$ and there are no dimension-specific zeros for $D > 4$. 

\item[Case IV:] $q  = 0, p \ge 0$. Finally, we consider the case $q = 0$. Here the coefficient of $x^2$ is
\be 
(-1)^p 2^{ p} \left[p(p-1) + \frac{p(p-1)}{4} +  p^2 + 1 \right] + 4p^2 \, ,
\ee
which vanishes for $p=1$ but otherwise has the sign $(-1)^p$. The coefficient of $x$ is
\be 
(-1)^p 2^{ p} \left[3p + 3  \right] + 4(1+2p) \, ,
\ee
which can easily be seen to have the correct sign. Finally, the constant term is now
\be 
4 (2^p) (-1)^p + 8
\ee
which has the correct behaviour. Therefore, for the case $q = 0$ and $p \ge 0$, the coefficients of the polynomial are always of the sign $(-1)^p$ and there are no dimension-specific zeros for $D > 4$. 
\end{description}

We have therefore shown that for all non-negative values of $p$ and $q$, $\beta_{p,q}$ when written in terms of $x = D - 4$ gives rise to a polynomial in $x$ that possesses all coefficients of the same sign. Therefore, by Descartes' rule of signs we can conclude that $\beta_{p,q}$ (excluding the cases $(p,q) = (1,0), (0,1)$ and $(1,1)$ where $\beta_{p,q}$ trivially vanishes), has no zeroes for positive real $x$. This then tells us that $\beta_{p,q}$ cannot have any dimension specific zeroes. Therefore, the expressions for the quasi-topological Lagrangians presented in the main text can be seen to hold at all orders and in all dimensions $D > 4$. 

%\renewcommand{\leftmark}{\MakeUppercase{Bibliography}}
%\phantomsection
\bibliographystyle{JHEP}
\bibliography{Gravities}
%\label{biblio}

\end{document}